\documentclass[article]{aa}

\usepackage{graphicx}
\usepackage[dvipsnames]{xcolor}
\usepackage{float}
\usepackage{longtable,booktabs}
\usepackage{amsmath}	
\usepackage{multirow}
\usepackage{url}
\usepackage{ae,aecompl}
\usepackage[flushleft]{threeparttable}
\usepackage{wasysym} 
\usepackage{scalerel} 
\usepackage{verbatim} 
\usepackage{listings}
\usepackage{txfonts}
\usepackage{natbib,twoopt}
\usepackage{xspace}
\usepackage{balance}
\usepackage{upgreek}
\usepackage{array}
\usepackage[normalem]{ulem}
\usepackage{soul}
\usepackage{outlines}
\usepackage{makecell}
\usepackage{siunitx}
\usepackage[colorlinks=true, allcolors=blue]{hyperref}
\bibpunct{(}{)}{;}{a}{}{,}
\DeclareGraphicsExtensions{.eps,.ps,.pdf}
\makeatletter
\renewcommand*\aa@pageof{, page \thepage{} of \pageref*{LastPage}}
\makeatother

\definecolor{grey}{rgb}{0.4,0.5,0.6}
\definecolor{brown}{rgb}{0.65,0.16,0.16}
\definecolor{darkgreen}{rgb}{0.0,0.45,0.0}
\definecolor{darkorange}{rgb}{0.9,0.2,0.0}

\begin{document}

   \title{Compact groups of galaxies in GAMA}

   \subtitle{Probing the densest minor systems at intermediate redshifts}

   \author{ A. Zandivarez\orcid{0000-0003-1022-1547}\inst{1,2}\fnmsep\thanks{\email{ariel.zandivarez@unc.edu.ar}}\and
E. Díaz-Giménez\orcid{0000-0001-5374-4810}\inst{1,2}    
          \and
          A. Taverna\orcid{0000-0003-1864-005X}\inst{1,2,3}
          \and
          F. Rodriguez\orcid{0000-0002-2039-4372}\inst{1,2}
          \and 
          M. Merchán\orcid{0000-0002-8085-7754}\inst{1,2}
          }

   \institute{Universidad Nacional de C\'ordoba (UNC). Observatorio Astron\'omico de C\'ordoba (OAC). C\'ordoba, Argentina
         \and
             CONICET. Instituto de Astronom\'ia Te\'orica y Experimental (IATE). Laprida 854, X5000BGR, C\'ordoba, Argentina
        \and
            Instituto de Astronomía, UNAM, Apdo. Postal 106, Ensenada 22800, B.C., México
             }

   \date{Received XXX XX 2023 / Accepted XXX XX 2024}

 
  \abstract
   {Over the years, several compact group catalogues have been constructed using different methods, but most of them are not deep enough to go beyond the very local universe with a high level of redshift completeness.}
   {We proposed to build a statistically reliable sample of compact groups to study the influence of its inner extreme environment at intermediate redshifts.}
   {We adopted the Galaxy And Mass Assembly redshift survey as a parent galaxy catalogue, complemented with a small sample of galaxies from the Sloan Digital Sky Survey, to identify compact groups using Hickson-like criteria. We explored the parameter space to perform several identifications to build samples with different characteristics. Particularly, we reduced the maximum galaxy separation in the line-of-sight to 500 km $s^{-1}$ and we implemented different magnitude ranges to define membership: 3, 2 or 1 magnitude difference between the brightest galaxy and the other members, and no restriction at all. For comparison, we used control samples extracted from a catalogue of loose groups to contrast properties with the compact groups.}
   {We build five considerably large compact group samples,  ranging from more than 400 up to roughly 2400 systems, and maximum redshifts from 0.2 to 0.4, depending on the selected parameters. The overall properties of each sample are in agreement with previous findings. Moreover, 
  there is a tendency of compact groups to have a larger fraction of quenched galaxies than control loose groups, mainly for low stellar mass galaxies in compact groups with small crossing times. In addition, $\sim 45\%$ of compact groups are embedded in loose galaxy systems and display the highest compactness, lowest crossing times and brightest first-ranked galaxies compared to compact groups considered non-embedded or isolated. There is almost no evolution of compact group properties with redshift.}
   {Our results confirm previous findings that postulate compact groups as one of the suitable places to study the suppression of the star formation rate in galaxies primarily due to galaxy interactions. These new Hickson-like compact group samples will be valuable to deepen the analysis of these peculiar galaxy systems in a redshift regime poorly explored so far.}


   \maketitle
%

\section{Introduction}
Since the discovery of the first compact groups (hereafter, CGs) of galaxies \citep{Stephan1877,Seyfert48} these systems have drawn the attention of scientists as a privileged setting to study the formation and evolution of galaxies in extreme environments. Its apparent physical nature described by a few galaxies confined in a small space generated the idea that the interactions between them should have been common and, therefore, their physical properties should be a reflection of that past. Hence, the need for large samples of these systems to allow for further study of these hypotheses became evident. 

The subsequent efforts made during the seventies by the study of the CGs of compact galaxies originally discovered by Shakhbazyan in 1957 (e.g., \citealt{Robinson&Wampler73,Shakhbazyan73,shakh2,shakh3}), triggered a systematic search of CGs leading to the first attempt to construct a CG catalogue made by \cite{Rose77}. 
These pioneering studies laid the foundations for what would be the construction of the most famous catalogue of compact groups in extragalactic literature, the one developed by \cite{Hickson82}. 
They visually selected 100 CGs in the plane of the sky from the Palomar Sky Survey prints based on galaxy associations with four or more bright galaxies with high mean surface brightness and without nearby galaxy companions. These selection requirements are known as population, compactness, and isolation, respectively.  
\cite{Hickson92} added redshift information which allowed them to introduce a new criterion in the line of sight (velocity concordance criterion) reducing the sample to 92 confirmed CGs in redshift space with three or more galaxy members. 
The overall procedure, which involves these four criteria is named as Hickson's criteria.

During the last thirty years, several attempts have been made to construct new CG samples. 
There have been two main procedures to identify CGs in galaxy surveys: the first was using a Hickson-like algorithm (e.g., \citealt{Prandoni+94,Iovino+02,Iovino+03,Lee+04,decarvalho+05,pompei+06,McConnachie+09,pompei+12,DiazGimenez+12,sohn+15,DiazGimenez+18,zheng+20,soto+22,zandivarez+22}), while the second option used a Friend-of-Friends (FoF) algorithm (i.e., a percolation algorithm that links galaxies in the transversal and radial directions in the sky) specially tuned to identify very dense galaxy systems with low membership (e.g., \citealt{Barton96,Focardi&Kelm02,deng+08,Hernandez&Mendes15,sohn+16}). 
As special cases, we can mention that a few studies have attempted to classify loose groups according to their compactness (e.g., \citealt{Zandivarez03,zheng+22}). 
Many of these previous works have been used to study several physical and dynamic properties of CGs.
Among the different observational works on CGs, we can mention as an example of the variety of topics, those related to their surrounding environment \citep{mendel+11,DiazGimenez+15,lee+17,zheng+21,taverna+23}, gravitational lensing around them \citep{chalela+17,chalela+18}, the gas depletion in galaxy members \citep{martinez+10,Bitsakis+15,lenkic+16,jones+19} and their path to the suppression of star formation \citep{walker+12,alatalo+15,Bitsakis+16,lisenfeld+17}, galaxy physical properties such as luminosity, age and metallicity and their comparison with other systems \citep{proctor+04,coenda+12,martinez+13,zheng+21,zandivarez+22}, different types of galaxy populations, morphology and stellar content \citep{Kelm&Focardi04,Plauchu-Frayn+12,zucker+16,moura+20}, and their faint galaxy companions \citep{campos+04,krusch+06,darocha+11,konstantopoulos+13, Zandivarez+14}.  

All these works were carried out either with Hickson's original sample or with subsequent catalogues which have in common that they are mainly confined to the local universe. 
All catalogues created with the FoF algorithm, and most of those that used a Hickson-like algorithms have constructed samples with the CGs located on average at $z < 0.1$. 
In contrast, only a few samples have reached a $z \sim 0.13$ on median \citep{Lee+04,decarvalho+05,pompei+06,pompei+12}, while others that have the potential to go beyond the local universe \citep{McConnachie+09} have very few redshift measurements. 
Only a few works have managed to venture into confirming CGs at redshifts close to 0.3 (three CGs by \citealt{gutierrez+11}), 1 (one CG by \citealt{gordon+23})  or beyond 2 (one CG each by \citealt{shih+15} and \citealt{nielsen+22}). 
Therefore, to deepen our understanding of the evolution of CGs, it is necessary to have catalogues with a considerably larger number of these systems beyond the local universe to perform reliable statistical studies of their main physical characteristics. 

To achieve this goal, it is essential to have galaxy surveys whose apparent magnitude limit is large enough while the redshift completeness is high. One of the redshift surveys that accomplish these characteristics is the Galaxy and Mass Assembly (hereafter, GAMA) spectroscopic survey \citep{gama}. This catalogue is almost two apparent magnitudes fainter than the Sloan Digital Sky Survey (SDSS) with a very high redshift completeness. 
Hence, in this work, we use the GAMA survey to build statistically fair CG samples that probe mainly the range of intermediate redshifts ($0.1 \lesssim z \lesssim 0.4$), a regime poorly explored in the literature.     

The layout of this work is as follows. In Sect.~\ref{sec:samples} we present the parent galaxy catalogue used in this work and the adopted sample loose groups to use as control samples, while in Sect.~\ref{sec:cgs}, we describe the identification of CGs performed on the parent survey. In Sect.~\ref{sec:props}, we analyse several CG properties and their comparison with the sample of loose groups. We summarise and discuss our results in Sect.~\ref{sec:conclusions}.  

\section{The galaxy and loose group samples}
\label{sec:samples}

\subsection{The parent galaxy catalogue}
\label{sec:samples_gama}
The main galaxy survey adopted in this work is the GAMA\footnote{Available in \url{http://gama-survey.org/}} spectroscopic survey \citep{gama, gama2}. 
The GAMA survey is an optical spectroscopic redshift survey designed to provide the best possible dataset for low to intermediate redshift galaxies with very high spectroscopic completeness over a volume suitable for performing fair statistical studies. The GAMA survey builds upon the two-degree field galaxy redshift survey \citep{2df1} and the Sloan Digital Sky Survey (SDSS; \citealt{york+00}). The survey extends over five discontiguous fields in the sky, where three of them lie in the equatorial part of the SDSS main area. 

We adopt only those GAMA galaxies belonging to the equatorial region's cones \citep{baldry+10}. The cones selected are those named in the GAMA literature as ${\bf G09}$ (RA:[129$^\circ$;141$^\circ$] - DEC:[-2$^\circ$;3$^\circ$]), ${\bf G12}$ (RA:[174$^\circ$;186$^\circ$] - DEC:[-3$^\circ$;2$^\circ$]) and ${\bf G15}$ (RA:[211.5$^\circ$;223.5$^\circ$] - DEC:[-2$^\circ$;3$^\circ$]). 
Each equatorial survey field covers a region of 5 × 12 sq degrees producing a total angular coverage of 180 deg$^2$. For these fields, redshifts have been obtained mainly by the GAMA team using the AAOmega facility at the Anglo-Australian Telescope. 
We select those galaxies that fulfil the redshift quality parameter $NQ\ge3$ (i.e., with a probability larger than 90\% to be correct, \citealt{liske+15}), and observed colour $g-r\leq 3$ and heliocentric redshift $> 0.003$, to avoid stars. We use the extinction-corrected apparent $r$ SDSS model mags in the AB system and transform the redshift into the CMB rest frame.
We adopt an apparent magnitude limit of 19.7 that ensures redshift completeness larger than 98\%. We also limit our sample to galaxies with redshifts lower than 0.5. We set this limit because there are very few galaxies beyond that distance, and it is also the maximum redshift for applying the k-correction algorithm \citep{Chilingarian+12} used in this work.

We complement the GAMA equatorial cones with galaxies with apparent magnitudes $r \leq 17.77$ using the galaxy redshift survey extracted from the SDSS Data Release 16 \citep{dr16} in the main contiguous area of the Legacy Survey as well as the compilation of \cite{tempel17}\footnote{Available in \url{http://cosmodb.to.ee}} made for the SDSS Data Release 12 \citep{DR12a,DR12b} and some corrections\footnote{They added already existent redshift estimations from other catalogues and discarded objects misclassified as galaxies.} made by \cite{DiazGimenez+18}. This galaxy sample comprises $565 \, 224$ galaxies with observer-frame model magnitudes $r \leq 17.77$ and observer-frame colour $g - r \leq 3$ to avoid stars. After performing a cross-match between GAMA and SDSS we obtained a sample of $12 \, 409$ galaxies with $r \leq 17.77$ to be added to the GAMA data.
Therefore, the final sample in the GAMA equatorial cones comprises $157\,634$ galaxies.

To estimate the galaxy rest-frame absolute magnitudes, k-corrections are computed using the code developed by \cite{Chilingarian+12}. The cosmology used here is the one obtained by the \cite{Planck+14}: $\Omega_m=0.31$ (matter density parameter), $h=0.67$ (dimensionless $z=0$ Hubble constant) and  ${\sigma}_8=0.83$ (standard deviation of the power spectrum on the scale of $8\,h^{-1}\,\rm Mpc$).

Finally, for the GAMA galaxies ($\sim 92\%$ of the final sample) we use the stellar masses estimated by \cite{taylor+11}. The Synthesis of Stellar Population (SPP) model was developed by \cite{ssp_bc03} with an initial mass function (IMF) of \cite{chabrier+03} and a dust curve by \cite{calzetti+00}. On the other hand, the specific star formation rates (sSFR) were derived by \cite{davies+16} using the MAGPHYS code \citep{magphys}. This code provides an estimate of the galaxy sSFR using SSP of \cite{ssp_bc03}, with a \cite{chabrier+03} IMF and an angle-averaged attenuation model of \cite{charlot+00}. 
For the remaining sample of SDSS galaxies, the stellar masses and sSFRs were extracted from the MPA-JHU public catalogue\footnote{Available in \url{http://www.mpa-garching.mpg.de/SDSS/}} \citep{kauffmann+03,brinchmann+04,tremonti+04,salim+07}.

\begin{figure}
   \centering
   \includegraphics[width=0.5\textwidth]{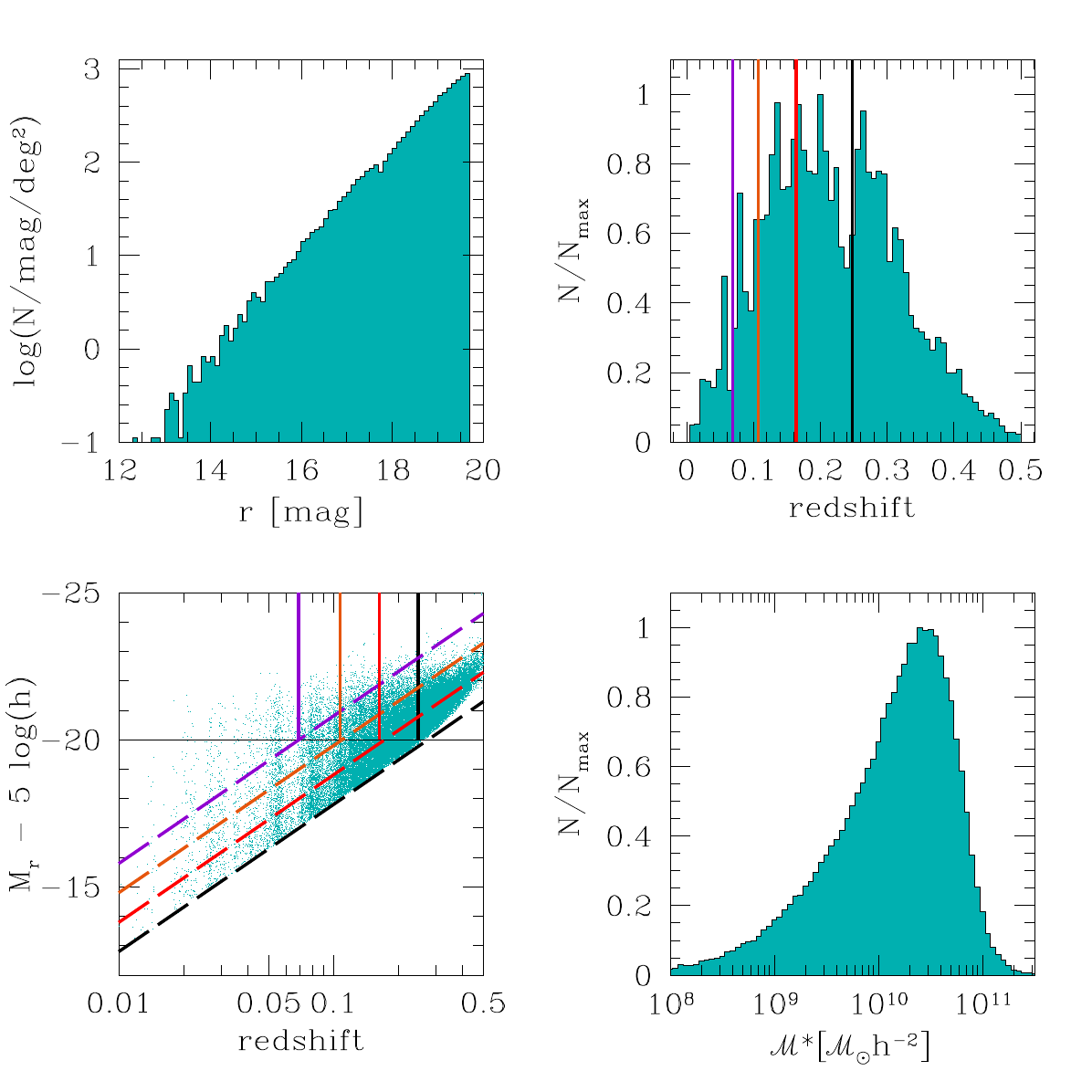}
      \caption{Distribution of properties for GAMA galaxies in the main galaxy sample used in this work. In the different panels, we display the galaxy number counts (top left), redshift distribution (top right), absolute magnitude in the r-band vs redshift (bottom left) and the galaxy stellar mass distribution (bottom right). Dashed lines in the bottom left panel show different apparent magnitude limits: 19.7 (black, main catalogue limit) as well as 18.7, 17.7 and 16.7  (red, orange and violet). Vertical lines (top right and bottom left panels) represent the redshift limits to define volume-limited samples used in Sect.~\ref{sec:zevol} and \ref{sec:tr}.
              }
         \label{fig:1}
\end{figure}
\begin{figure}
   \centering
   \includegraphics[width=0.5\textwidth]{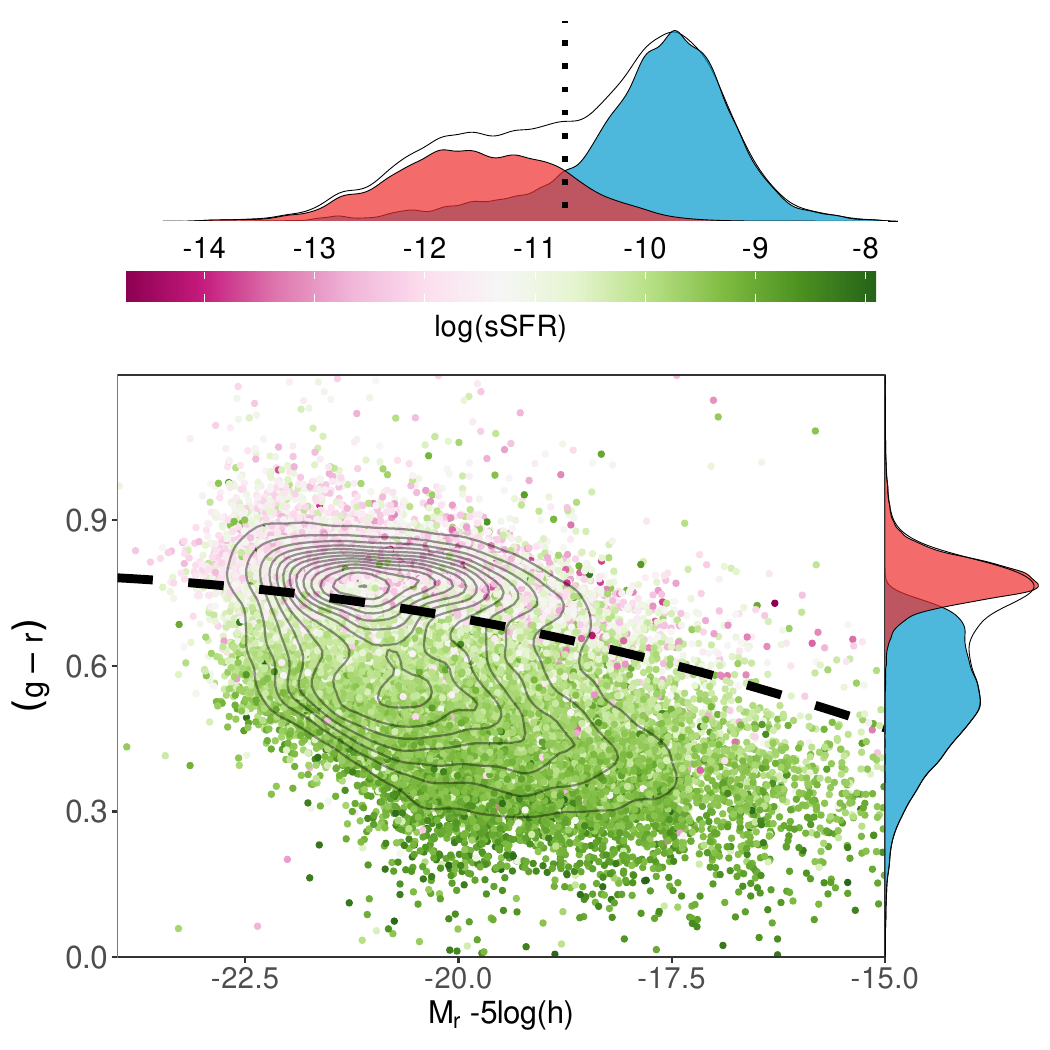}
      \caption{Galaxy colour-magnitude diagram for GAMA galaxies. The logarithm of the specific star formation rate for each galaxy is displayed according to the upper colour bar. The dashed black line in the main panel indicates the empirical law to separate between red and blue galaxies (see Section~\ref{sec:samples}). The right marginal plot shows the colour distributions for galaxies classified as red (above the dashed line) and blue (below the dashed line) galaxies. The upper distributions show the specific star formation rate for galaxies classified as red and blue following the empirical law. The vertical dotted line indicates the specific star formation limit (-10.7) adopted to split the galaxy sample into quenched galaxies (to the left) and star-forming galaxies (to the right).}
         \label{fig:2}
\end{figure}

Some of the properties of the final galaxy sample are shown in Fig.~\ref{fig:1}. 
The top left panel displays the galaxy number counts for the galaxy sample (apparent magnitude cut of $19.7$ in the SDSS r-band). 
The top right panel shows the redshift distribution of the galaxy sample, reaching a maximum of $0.5$ with a median of $0.21$.
The r-band absolute magnitude as a function of redshift is shown in the bottom left panel. In this panel, the lower envelope black dashed line indicates the magnitude-redshift relation for the adopted apparent magnitude limit of the catalogue, the other dashed lines show one (red), two (orange) or three (purple) magnitudes brighter than the magnitude limit of the survey. These three limits will be relevant in identifying CGs in Sect.~\ref{sec:cgs}. 
Finally, the bottom right shows the galaxy stellar mass distribution with a median value of $ 1.66 \times 10^{10} \ {\cal M}_{\odot} \ h^{-2}$.

We also classify galaxies into distinct galaxy populations. 
Firstly, we consider the bimodal behaviour of the rest-frame galaxy colour $g-r$ to differentiate between red and blue galaxies. Since the $g-r$ distribution is a function of the absolute magnitude of the galaxy, we follow a procedure similar to that performed by \cite{zm+11} and recently applied to SDSS DR16 galaxies in \cite{taverna+23}.   
Briefly, we divide the whole range of r-band absolute magnitudes into several bins, and we fit two Gaussian functions to the $g-r$ colour distribution of galaxies within each absolute magnitude bin\footnote{This procedure is performed using Gaussian mixture models provided by the {\it Mclust package} of R software \citep{mclust+16}.}. 
Then, using the colour value of the intersection of these Gaussian functions for each bin, we fit a 2-degree polynomial empirical law as a function of the absolute magnitude. The best fit is $P(x)= -0.00274 x^2 - 0.14121 x - 1.02936$ with $x=M_r-5\log(h)$. Slight variations of these coefficients do not change the final statistical results of this work.
Figure~\ref{fig:2} shows the scatter plot of $g-r$ vs r-band absolute magnitude for galaxies in the main sample. The dashed line in this figure is the 2-degree polynomial function $P(x)$ used to differentiate between red (above the curve: $\sim 35\%$ of the sample) and blue (below the curve: $\sim 65\%$) galaxies. The $g-r$ distributions for the resulting red and blue galaxy populations are shown in the marginal distribution at the right of Fig.~\ref{fig:2}.

In the literature, the usage of galaxy colours and sSFR is common as clear indicators of the different evolution in terms of their ability to form stars (see for instance \citealt{weinmann+06} that analysed the impact of environment on galaxies classified according to both, galaxy colours and sSFR). Usually, it is expected that red galaxies are characterized by having low sSFR while blue galaxies typically display high sSFR. However, it is likely that due to a very strong extinction, some star-forming edge-on disc galaxies may appear red in the colour-magnitude diagram. Therefore, the sSFR is more suitable to isolate those galaxies with suppressed star formation (quenched) from those that seem to be currently forming stars. Hence, we also use the sSFR to separate our main galaxy sample into two galaxy populations. In the colour-magnitude diagram of Fig.\ref{fig:2} we have coloured galaxies according to their logarithmic sSFR value. At the top of the diagram, we display the total bimodal distribution for the sSFR of galaxies in the sample (black line) and the distributions obtained for the populations of red and blue galaxies classified above. Several limits have been proposed in the literature over the years to split the sSFR distribution, which goes from -11 \citep{wetzel+12,henriques+17,Ayromlou21} to -10.5 \citep{lacerna+22}. We select the intersecting value between the red and the blue sSFR distributions, -10.7, to separate quenched (lower values) and star-forming (higher values) galaxies. This limit agrees with \cite{brown+17} and \cite{Cora+18} which have also selected this value to differentiate between two galaxy populations for galaxies in the SDSS and the Multidark simulation \citep{klypin+16}, respectively. 

\begin{figure}
    \centering
    \includegraphics[width=0.45\textwidth]{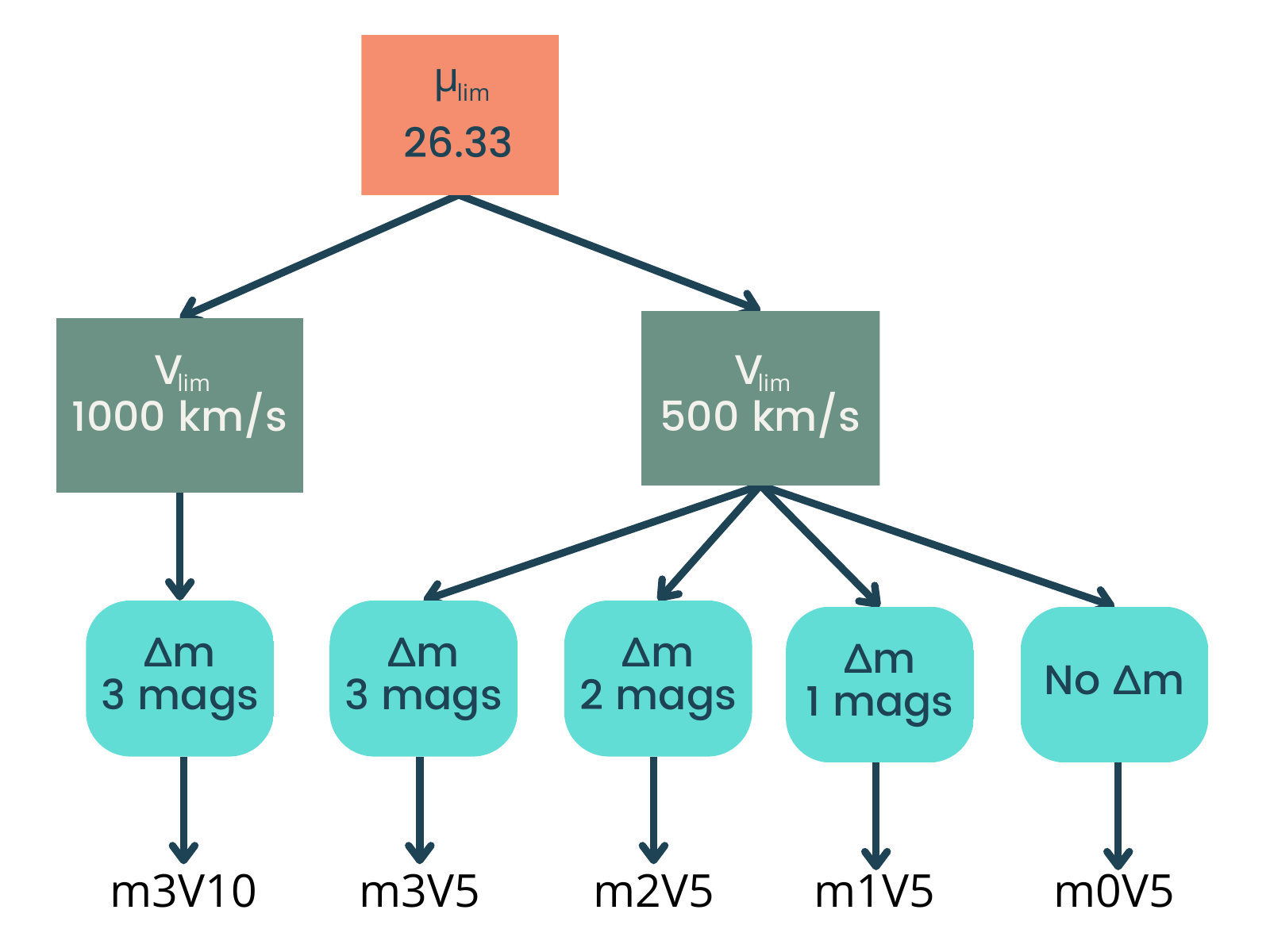}
    \caption{Flowchart displaying the identification parameters adopted to build five different CG samples from the GAMA survey.}
    \label{fig:scheme}
\end{figure}

\subsection{Loose Galaxy Groups}
\label{sec:LGs}
We extract loose galaxy groups from the above sample using the procedure described in \cite{rodriguez2020combining}. This method initially applies a FoF algorithm to identify gravitationally bound galaxy systems with at least one bright galaxy with an r-band absolute magnitude brighter than -19.5. Subsequently, a halo-based algorithm is applied \citep{yang+05,yang+07}. Using potential members of FoF galaxies in redshift space, the algorithm calculates a three-dimensional density contrast by determining a characteristic luminosity. The estimation procedure considers the incompleteness caused by the limiting magnitude of the observational catalogue. By abundance matching in the luminosity, we assign the mass of each group. We assume that the galaxies populate the DM halos according to a \cite*{navarro1997universal} profile. We calculate the three-dimensional density contrast using the assigned mass to associate the galaxies with the groups. With this final membership assignment, the procedure recalculates the characteristic luminosity and iterates until it converges. This method reliably identifies galaxy systems with both, low and high membership. For a detailed description of this algorithm and how it works, see \cite{rodriguez2020combining}. 
It is important to note that, while in this specific context, this sample of normal groups will be employed as control sample groups, in other studies, this procedure has been used to identify galaxy systems for different goals such as investigating scale relationships \citep{rodriguez+21}, estimating mass through weak lensing \citep{gonzalez+21}, and examining the influence of the environment on member galaxies \citep{alfaro+22,rodriguez+22,rodriguez-medrano+23}, among other applications. 

The sample of loose groups in GAMA comprises 94423 groups with one or more galaxy members and is made publicly available in this work. We include a brief description of the catalogue in Appendix~\ref{app_t}. Here, we select loose groups with three or more galaxy members. The final number of loose groups with three or more members is $6\,844$. 

\begin{table}
\caption{CG identification criteria and number of CG identified in each sample \label{tab:samples}} 
    \begin{center} 
    \tabcolsep=0.5pt
    \begin{tabular}{lcccccc}
         \cline{2-6}
         & {\tiny Population} & {\tiny BGG Flux} & {\tiny Compactness} & {\tiny Isolation} & {\tiny Velocity} \\
         \cline{2-6}
         &  $N \ge 3$ & $r_b < r_{blim}$ & $\mu \le \mu_{lim}$ & {\tiny $\Theta_I \ge 3\Theta_G$} & {\tiny $\Delta V \le V_{lim}$} \\
         &  $N \in \Delta m$ & {\tiny $r_{blim} = r_{lim}- \Delta m$} & & $N \in \Delta m$ & \\
        \cline{2-6}
         & $\Delta m$ & $r_{lim}$ ; $r_{blim}$ & $\mu_{lim}$ & $\Delta m$ & $V_{lim}$ & $N_{CG}$\\
        \hline
       Z+22  & 3 & 17.77 ; 14.77 & $26.33$ & 3 & $1000$ & 1412 \\ 
       m3V10  & 3 & 19.70 ; 16.70 & $26.33$ & 3 & $1000$ & $ \ \ 467$\\
       m3V5  & 3 & 19.70 ; 16.70 & $26.33$ & 3 & $ \ \ 500$ & $ \ \ 438$\\ 
       m2V5  & 2 & 19.70 ; 17.70 & $26.33$ & 2 & $ \ \ 500$ & $ \ \ 759$\\
       m1V5  & 1 & 19.70 ; 18.70 & $26.33$ & 1 & $ \ \ 500$ & $ \ \ 612$\\
       m0V5  & None & 19.70 ; 19.70 & $26.33$ & All & $ \ \ 500$ & 2356\\
       ZS+20  & None & 17.77 ; 17.77 & $26.00$ & All & $1000$ & 6144\\
       \hline
    \end{tabular}
    \end{center}
    \parbox{\hsize}{Notes: $N$ is the number of galaxy members; $r_b$ is the apparent magnitude of the brightest galaxy; $\Theta_G$ is the radius of the smallest circle that enclosed all galaxy members; $\Theta_I$ is the isolation radius for the nearest non-member galaxy within $\Delta m$; $N_{CG}$ is the total number of CGs in each sample. Acronyms for previous works: Z+22 \citep{zandivarez+22} and ZS+20 \citep{zheng+20}. Units: $\mu_{lim}$ is in mag ${\rm arcsec^{-2}}$, while $V_{lim}$ is in km $s^{-1}$}
\end{table}
\begin{figure*}
   \centering
   \includegraphics[width=0.8\textwidth]{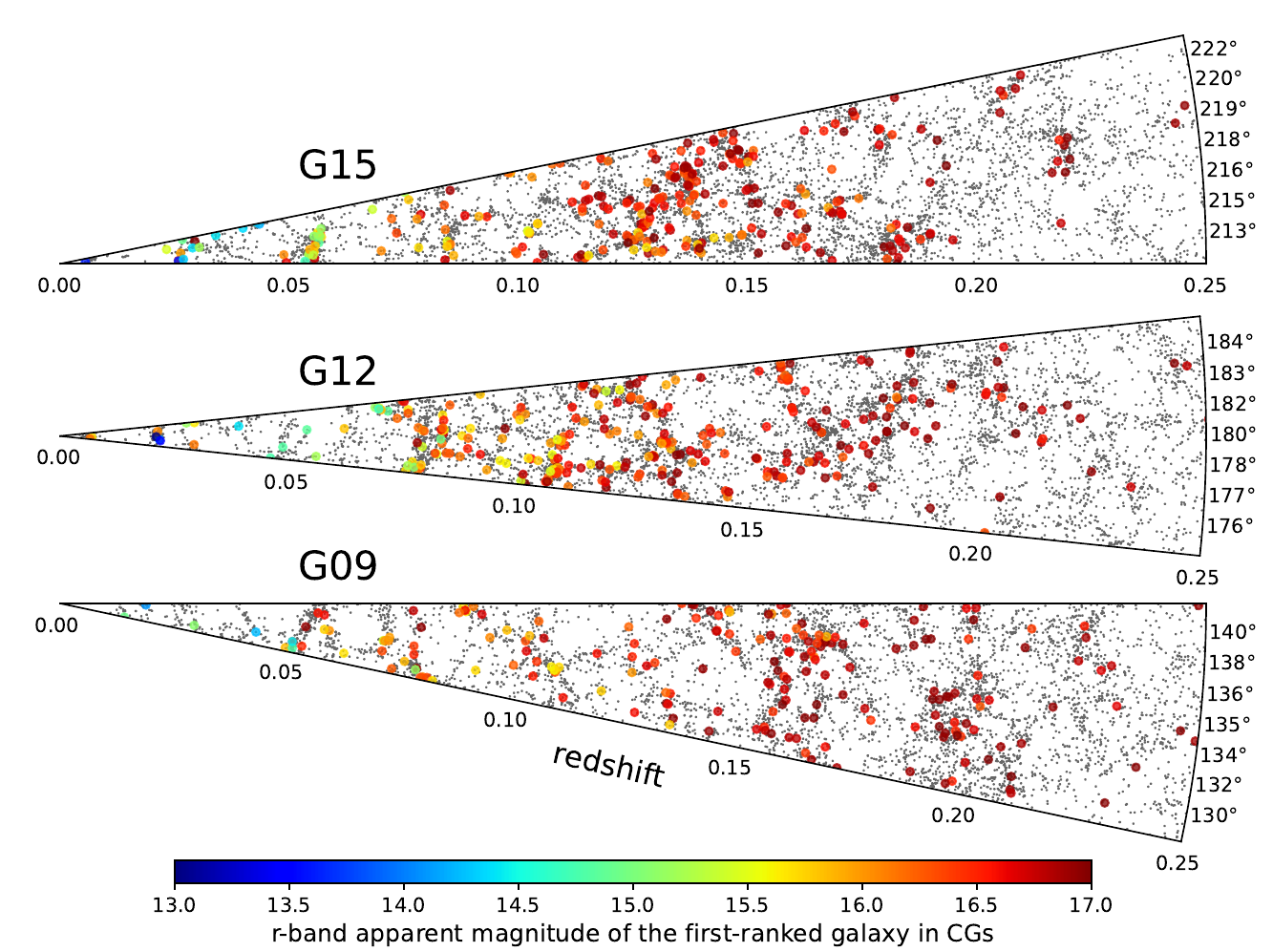}
      \caption{Spatial distribution of CGs identified in GAMA equatorial fields (big filled dots). In this case, the sample of CGs is the CG-m2V5, i.e., with members within a two-magnitude gap and $500\ km \ s^{-1}$ of maximum velocity separation from the CG centre. Each cone displays the CG positions using the right ascension and redshift in each field. The colour distribution shows the r-band apparent magnitude of the brightest galaxy in the CG. Black points are a random sample of galaxies ($\sim 20\%$) in the main survey. 
              }
         \label{fig:3}
\end{figure*}
\section{Compact Groups in GAMA}
\label{sec:cgs}
The samples of CGs are identified using the CG finding algorithm of \cite{DiazGimenez+18}. Following the definition devised by the original criteria of \cite{Hickson82} and \cite{Hickson92}, the algorithm looks for CGs that simultaneously satisfy constraints on membership, compactness of the group, relative isolation and velocity concordance, as well as flux limit of the brightest group galaxy (BGG) to ensure completeness (e.g., \citealt{Prandoni+94,DiazGimenez&Mamon10}). 
The overall criteria establish the following requirements: 
\begin{description}
    \item[\bf Population:] There are 3 or more galaxy bright members (in a $\Delta m$-magnitude range from the brightest)
    \item[\bf Flux limit of the BGG:] The BGG of the system has to be at least $\Delta m$ magnitudes brighter than the catalogue magnitude limit, $r_{\rm lim}$.
    \item[\bf Velocity concordance:] galaxy members are within $V_{lim}$
    from the median velocity of the system. 
    \item[\bf Compactness:] the surface brightness of the system in the $r$-band is less than $\mu_{lim}$ 
    \item[\bf Relative Isolation:] There are no other bright galaxies (in a $\Delta m$-mag range from the brightest, nor brighter) within three times the radius of the minimum circle that encloses the galaxy members.
\end{description}
where $\Delta m = 3$ magnitudes, $V_{lim} = 1\,000 \rm \, km\,s^{-1}$, and $\mu_{lim}= 26.33 \, \rm mag \, arcsec^{-2}$ (this is the 26 limit defined by \citealt{Hickson82} in the $R$ band but modified for the $r$ SDSS band).

These Hickson-like criteria described above are identical to ones used by \cite{zandivarez+22} in the SDSS DR16. By applying the same algorithm to the GAMA catalogue, we can nearly double the depth of the CG sample identified in the SDSS.

However, to exploit the redshift depth of the GAMA survey, we will perform a set of identifications changing some of the parameters of the Hickson-like algorithm and analyse the effect of these variations on the resulting samples. 

\subsection{Choosing identification parameters}

The parameters of the algorithm that can be tuned are: 
the limiting surface brightness ($\mu_{lim}$), the maximum velocity separation of galaxy members from the group centre ($V_{lim}$), and the magnitude gap ($\Delta m$) where the galaxy members are selected.

\begin{itemize}
    \item The compactness of the system is mostly defined by the setting of $\mu_{lim}$. The original value selected by \cite{Hickson82} was $26 \, \rm mag \, arcsec^{-2}$ in the R band. In the literature, some works directly used the same value regardless the photometric band (e.g., \citealt{McConnachie+09,zheng+20}). In contrast, others changed the Hickson value to match the photometric band of the parent galaxy catalogue (e.g., \citealt{Prandoni+94,Iovino+02,DiazGimenez+12,Taverna+16,DiazGimenez+18}), while others directly adopted lower limits to reduce contamination regardless the band (e.g., \citealt{Iovino+03,Lee+04,decarvalho+05}).

    \item \cite{Hickson92} analysed the difference in the radial velocity of the galaxy members of their CGs with the median velocity of the CG centre ($\Delta V$), and observed that the great majority of galaxy members lie within $\pm 1000 \ km \ s^{-1}$ from the centre. Therefore, he decided that this value should be set as the limit ($V_{lim}$) to select physically associated galaxies with the group. Several works have adopted this limit to identify CGs whether a Hickson or FoF type algorithm has been used (e.g., \citealt{Barton96,Focardi&Kelm02,DiazGimenez+12,sohn+15,sohn+16,DiazGimenez+18,zheng+20}). However, it must be considered that $1000 \ km \ s^{-1}$ in the line of sight is a fairly large distance ($\sim$ 10 Mpc h$^{-1}$) and therefore, the identification is susceptible to contamination. 
It is worth noting that roughly 77\% of galaxies in Hickson CGs differ less than 500 km/s from the median of the systems \citep{hickson+97}, while this percentage is nearly $\sim 95\%$ in the automatic catalogues produced by \cite{DiazGimenez+18, zheng+20,zandivarez+22}.  
Therefore, setting $V_{lim}$ to a lower value could help to avoid interlopers in the line of sight.

    \item Finally, regarding the $\Delta m$, while the original criterion is $\Delta m=3$, different authors in the literature have adopted different gaps. For instance, some authors adopted $\Delta m=2$ to reduce potential contamination at the expense of decreasing completeness \citep{decarvalho+05,Iovino+03}.  Others stated there is no point in using a magnitude gap when identifying CGs in redshift surveys. They argue that Hickson included this criterion to maximize the probability of having similar redshifts when selecting galaxies with similar magnitudes \citep{Barton96,zheng+20}. The $\Delta m$ also has an impact on the isolation criterion. 
As described above, the $\Delta m$ determines the galaxies inside the isolation region that can be considered intruders. 
Nevertheless, some authors have preferred to break that dependence  
because they considered the criterion too strict, which is detrimental to detecting systems in projection whose fainter members are near the selection limit. 
Most of those studies search for intruders in the isolation region within the magnitude range defined by the magnitude of the faintest members plus 0.35 or 0.5 mags \citep{Iovino+02,Iovino+03,decarvalho+05}. 
The selection of $\Delta m$ also affects the flux limit criterion. 
To ensure the completeness of the galaxy member selection within a $\Delta m$, the brightest galaxy must be brighter than $m_{lim}-\Delta m$. 
This criterion produces a CG sample that is homogeneous with redshift and suitable for unbiased statistical studies \citep{zheng+20}.
\end{itemize}
 
As mentioned above, different choices of parameters have been used in the literature. It is not the aim of this work to claim a set of parameters as the most adequate for identifying CGs but instead, we aim to produce several samples of CGs in GAMA that satisfy different restrictions that can be useful for different objectives. 

Therefore, in this work, we implement different combinations of free parameters. We keep $\mu_{lim}=26.33$, which is the compactness proposed by Hickson but adapted to the SDSS r-band, and vary the velocity limit (1000 and 500 km/s) and the magnitude gap (3, 2, 1 and none). In Fig.~\ref{fig:scheme} we show the combinations used to create different CG samples: {\tt m3V10}, {\tt m3V5}, {\tt m2V5}, {\tt m1V5}, and {\tt m0V5}.
The different magnitude gaps will allow us to obtain CG samples with different redshift distributions increasing the depth of the catalogue as $\Delta m$ decreases. We also include one 
catalogue without introducing any $\Delta m$. This is an inhomogeneous sample with redshift, but comparable with previous catalogues available in the literature.

The summary of the catalogues built in this work is outlined in Table~\ref{tab:samples}, as well as two previous catalogues that will be used for comparisons. As an example, Fig.~\ref{fig:3} shows the spatial distribution of one of the CG samples presented in this work ({\tt m2V5}) over-plotted on top of the spatial distribution of galaxies in GAMA (20\% of these galaxies).

\begin{figure*}
   \centering
   \includegraphics[width=0.49\textwidth]{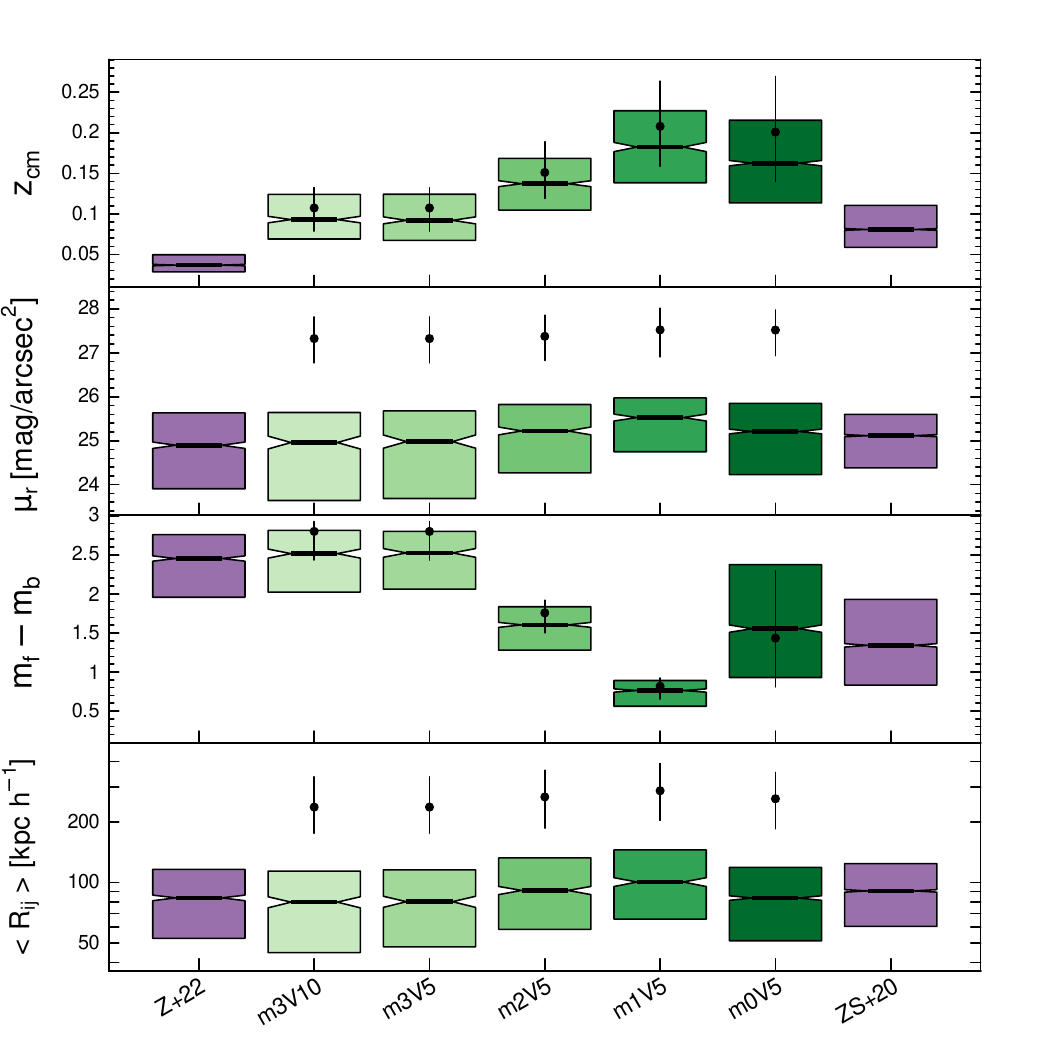}
   \includegraphics[width=0.49\textwidth]{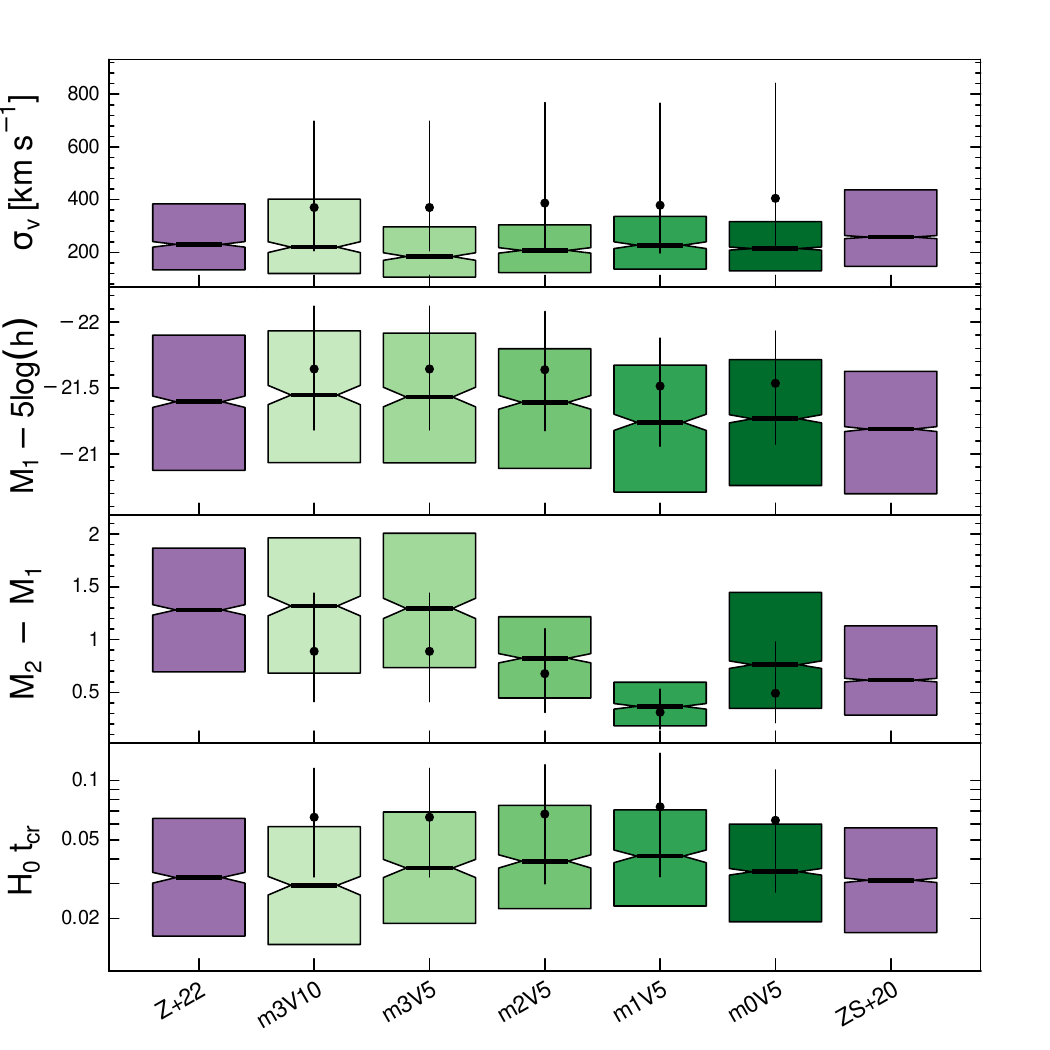}
      \caption{Boxplot diagrams of the properties of CGs. From top left to bottom right: median redshift of the galaxy members, radial velocity dispersion, surface brightness, absolute magnitude of the brightest galaxy, apparent magnitude difference between the brightest and faintest galaxy, absolute magnitude difference between the two brightest galaxies, median of the projected inter-galaxy distance, dimensionless crossing time. CGs identified in the GAMA survey are shown in green colours and labelled as m\#V\#, where the first index denotes the number of magnitudes used for the magnitude gap, while the second index corresponds to a hundredth of the limit for the velocity concordance criterion. 
      We include the boxplot diagrams of two previous SDSS CG samples in purple for comparison (see Table~\ref{tab:samples}) Dots with bars represent the median properties $\pm$ quartile ranges of control groups. 
      }
      \label{fig:4} 
\end{figure*}
\section{CGs properties analysis}
\label{sec:props}
\subsection{General distributions} 
In Fig.~\ref{fig:4} we show the boxplot diagrams of the distributions of properties. Green colours correspond to the five CG samples presented in this work for GAMA, while purple is adopted for previous SDSS CG samples.  
The properties are: 
\begin{itemize}
    \item $z_{\rm cm}$: bi-weighted median of redshifts of the galaxy members (eq. 5 from \citealt{Beers+90}). 
    \item $\mu_{\rm r}$: surface brightness of the group in the r-band computed inside the minimum circle that encloses the galaxy members. 
    \item $m_f - m_b$: difference in observer-frame apparent magnitude between the brightest and the faintest galaxy members. 
    \item $\langle R_{ij}\rangle$: median of the projected separations between galaxy members. 
    \item $\sigma_v$: radial velocity dispersion of the group computed with the gapper (n<10) or the biweight (n>10) estimators \citep{Beers+90}.
    \item $M_1-5log(h)$: rest frame absolute magnitude of the brightest galaxy of the group. 
    \item $M_2-M_1$: difference in absolute magnitudes between the brightest and the second brightest galaxy members.
    \item $H_0 \, t_{\rm cr}$: dimensionless crossing time of the groups computed as 
$100 \, {h} \pi\langle R_{ij} \rangle /(2 \sqrt{3}\sigma_v)$.
\end{itemize}
Some of these properties are directly linked to the parameters used to perform the identification and their variations reflect the change in the selection criteria ($\mu_{\rm r}$, $m_f - m_b$ and $M_2-M_1$). All these properties as well as the angular coordinates of the CG centres presented in this work are made publicly available. Appendix~\ref{app_t} includes the description of the released tables. 
 
For comparisons, we use the sample of loose groups described in Sect.~\ref{sec:LGs} as a control sample. Dots with error bars in Fig.~\ref{fig:4} display the behaviour of control samples (loose groups).  
Since CG properties are computed using galaxies in a given range of magnitudes, we also impose the magnitude range to select the galaxies in loose groups to measure the group properties and the flux limit on the brightest galaxy. 
We only work with those loose groups with 3 or more members within a given magnitude range from the brightest. 
Therefore, we have a control sample of loose groups for each CG sample. The control sample for {\tt m3V10} comprises 705 systems, while those with $V_{lim}= 500 \ km \ s^{-1}$ have 1899, 1752, 721 and 5523 systems for magnitude ranges of 1, 2 and 3 mags and no gap, respectively.

In addition, we compare our CG catalogues with previous CG catalogues available in the literature. \cite{zheng+20}\footnote{Available in \url{https://cdsarc.cds.unistra.fr/viz-bin/cat/J/ApJS/246/12}} presented a CG catalogue in the SDSS DR12 ($r_{\rm lim}=17.77$). In this work, the authors identified CGs with a Hickson-like algorithm with $\mu_{\rm lim}= 26 \, \rm mag \, arcsec^2$, $V_{\rm lim}=1000 \, \rm km/s$ and without imposing a magnitude gap. 
\cite{zandivarez+22}\footnote{Available in \url{https://cdsarc.cds.unistra.fr/viz-bin/cat/J/MNRAS/514/1231}}  
produced a CG catalogue in the SDSS DR16 ($r_{\rm lim}=17.77$) using a Hickson-like algorithm with $\mu_{\rm lim}= 26.33 \, \rm mag \, arcsec^2$, $V_{\rm lim}=1000 \, \rm km/s$ and a 3-magnitude gap.

\begin{figure*}
   \centering
   \includegraphics[width=0.49\textwidth]{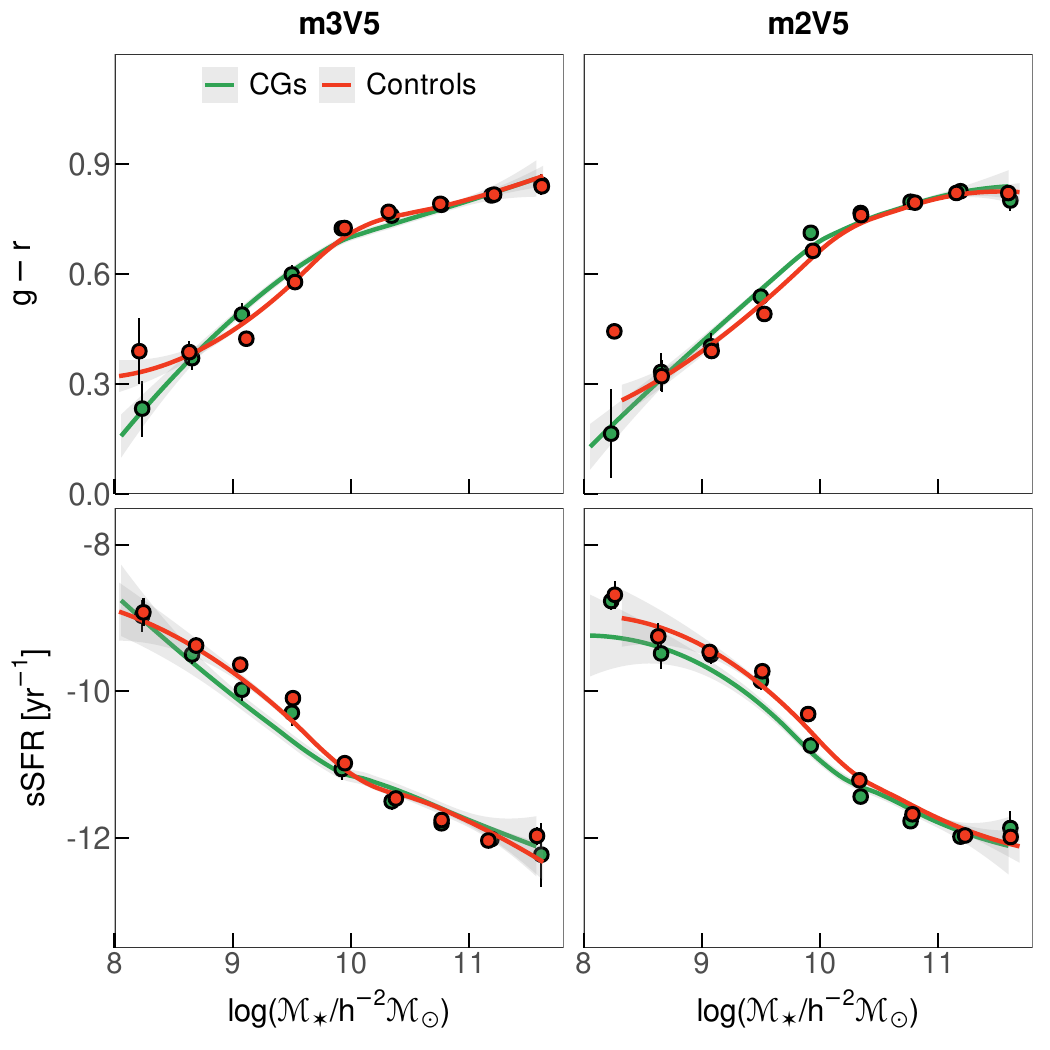}
   \includegraphics[trim={1.6cm 0 0 0}, clip, width=0.446\textwidth]{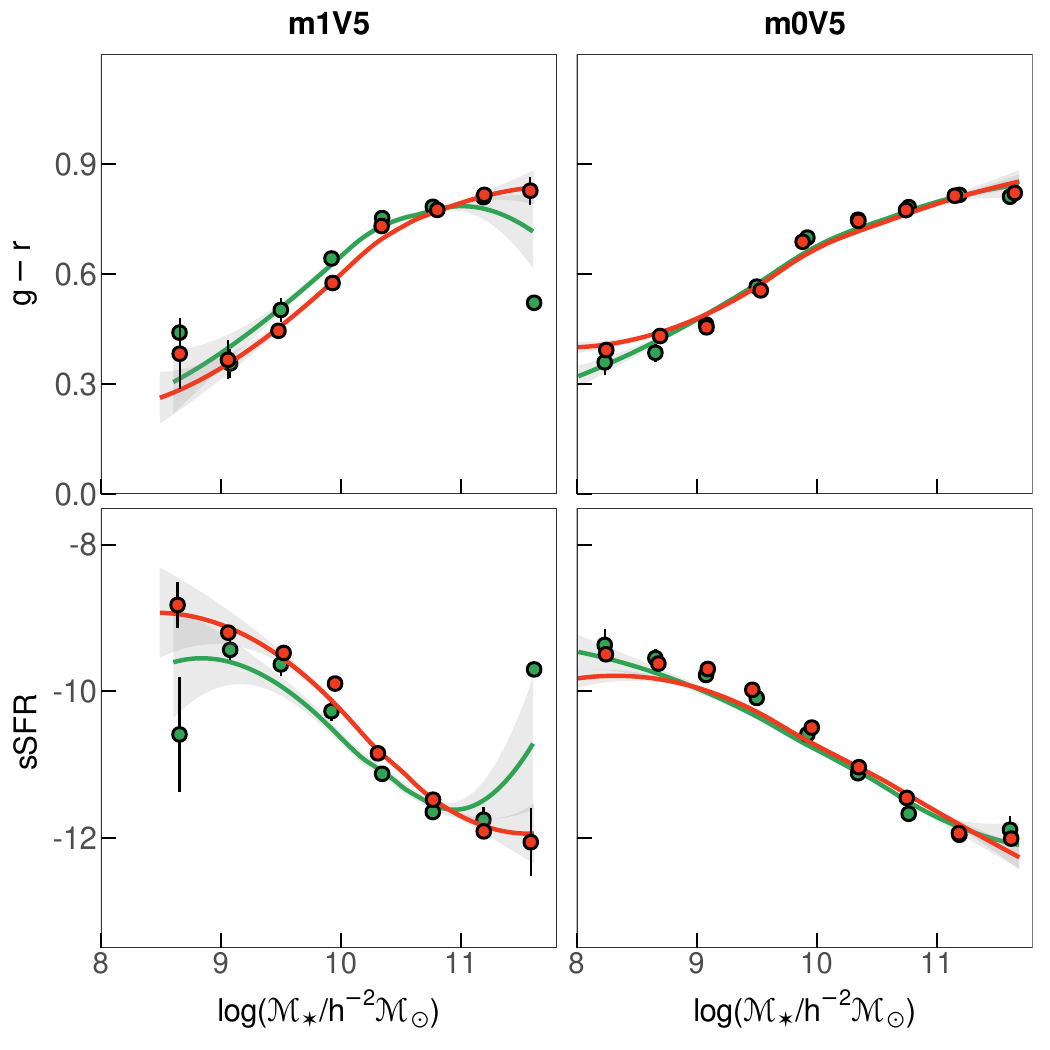}
   \includegraphics[width=0.49\textwidth]{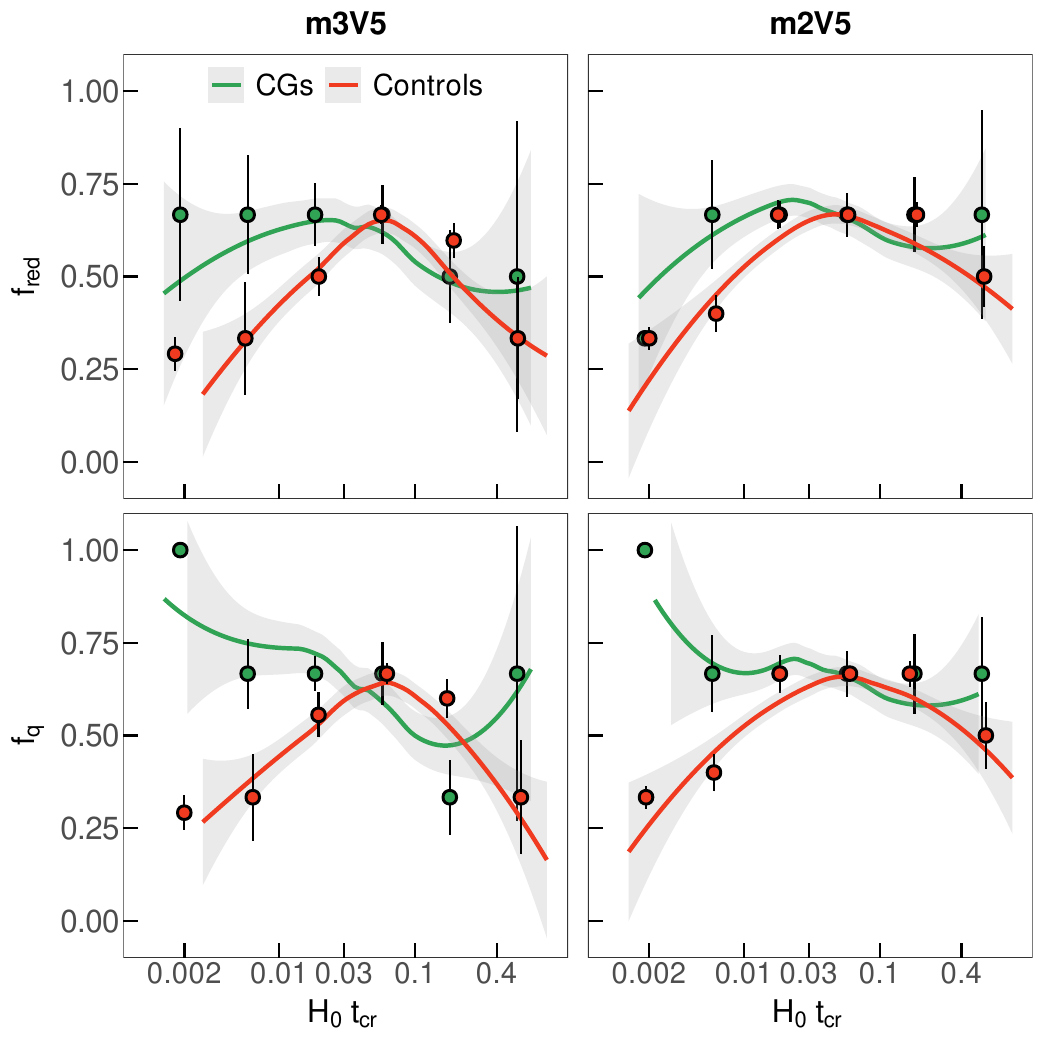}
   \includegraphics[trim={2cm 0 0 0}, clip, width=0.435\textwidth]{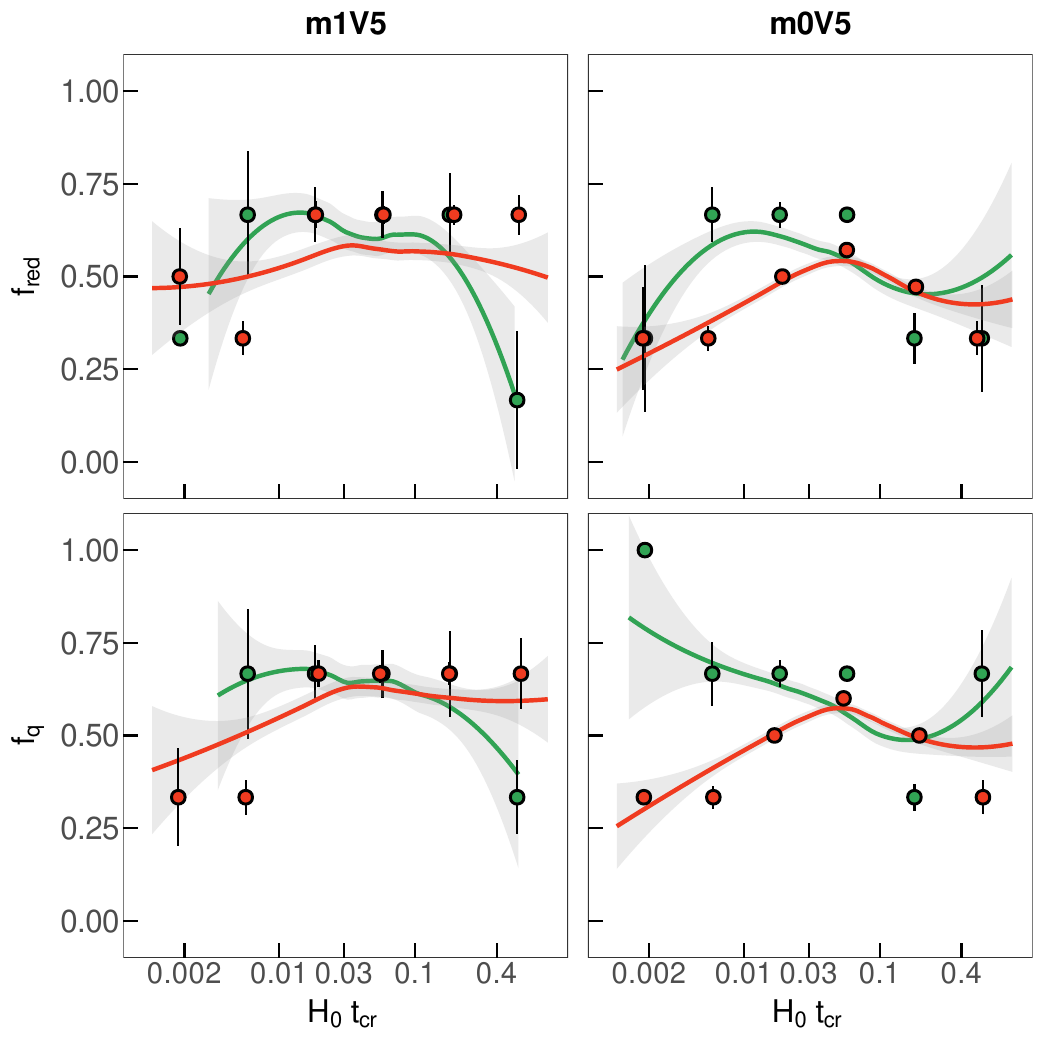}
      \caption{Upper plots: the distribution of the specific star formation rate and colour as a function of galaxy stellar masses for galaxy members in systems. Lower plots: the distribution of the fraction of red and quenched galaxies in systems as a function of their dimensionless crossing-time. In all panels, dots represent the median values per bin while error bars are the 95\% confidence interval for the median. Solid lines are the polynomial fitting to the scatter plot with the LOESS method and the grey region is the 95\% confidence interval. In each column, the comparison is made between CGs (green) and their control sample (red). Each sample have $V_{lim}=500 \ km \ s^{-1}$.     
       }
         \label{fig:6}
\end{figure*}

Comparing Z+22 with {\tt m3V10}, the only difference in the CG finder is the flux limit of the brightest galaxy since the apparent magnitude limits of the source catalogues are different ($r_{\rm lim}=17.77$ in SDSS DR16 vs. $r_{\rm lim}=19.70$ in GAMA). {\tt m3V10} comprises CGs with its brightest galaxy almost 2 magnitudes fainter than Z+22, therefore this sample goes deeper in redshift with a median of $0.09$ compared to the median redshift of $0.03$ for Z+22. All the other CG properties shown in Fig.~\ref{fig:4} are similar between these two catalogues (the notches of the medians overlap). 

The effect of changing the velocity limit from $1000 \, \rm km/s$ to $500 \, \rm km/s$ can be evaluated in the comparison between {\tt m3V10} and {\tt m3V5}. As mentioned above for other automatic catalogues, most of the galaxies in {\tt m3V10} are within $500$ km/s from the group centre, therefore the number of groups does not change significantly (from 467 to 438, see Table~\ref{tab:samples}). There is a slight decrease in the velocity dispersion in the {\tt m3V5} as a reflection of a smaller $V_{\rm lim}$ that has eliminated potential interlopers in the CGs, while all the other independent properties remain indistinguishable. 
 
As the magnitude gap changes, the flux limit of the brightest galaxy varies.
Then, moving from {\tt m3V5} to {\tt m2V5} to {\tt m1V5}, we include groups with the brightest galaxies 1 and 2 magnitudes fainter. Therefore, the depth of the samples increases as the magnitude gap decreases. Also, a smaller magnitude gap means we are selecting groups with more similar galaxies.  
The number of {\tt m2V5} almost doubles the {\tt m3V5} (Table~\ref{tab:samples}), and it decreases again in {\tt m1V5}. 
Moving from $\Delta m =3$ to 1, there is a small increase in the projected separation between galaxies and the velocity dispersion, while a small decrease is observed in the magnitude of the brightest galaxies. We will deepen the analyses of the possible dependence of the properties with redshift in the following sections.

Sample {\tt m0V5} is the most numerous but also inhomogeneous sample. There is no limit in the flux of the brightest galaxy nor a fixed magnitude gap to look for its neighbours. The latter means this sample is the most restrictive regarding isolation since it does not allow any galaxy to contaminate the isolation disk. 
Still, its properties do not differ significantly from the other samples, especially with the {\tt m2V5}. This sample is useful to compare with the sample of ZS+20 which was identified similarly but in the SDSS DR12 catalogue. The fainter magnitude limit of the GAMA catalogue allows us to go farther in redshift space, while the smaller $V_{\rm lim}$ used in this work produces CGs with smaller velocity dispersions, as noticed above. 
There are also small differences in the absolute magnitude of the brightest galaxy, and the magnitude gap between the two brightest galaxies. The larger gap is a consequence of the possibility in GAMA to go 2 magnitudes beyond the SDSS (allowing the possibility of measuring larger gaps), while the brighter absolute magnitude of the first-ranked galaxy in GAMA is due to a deeper sample (larger redshifts). 

\begin{figure}
   \centering
   \includegraphics[width=0.49\textwidth]{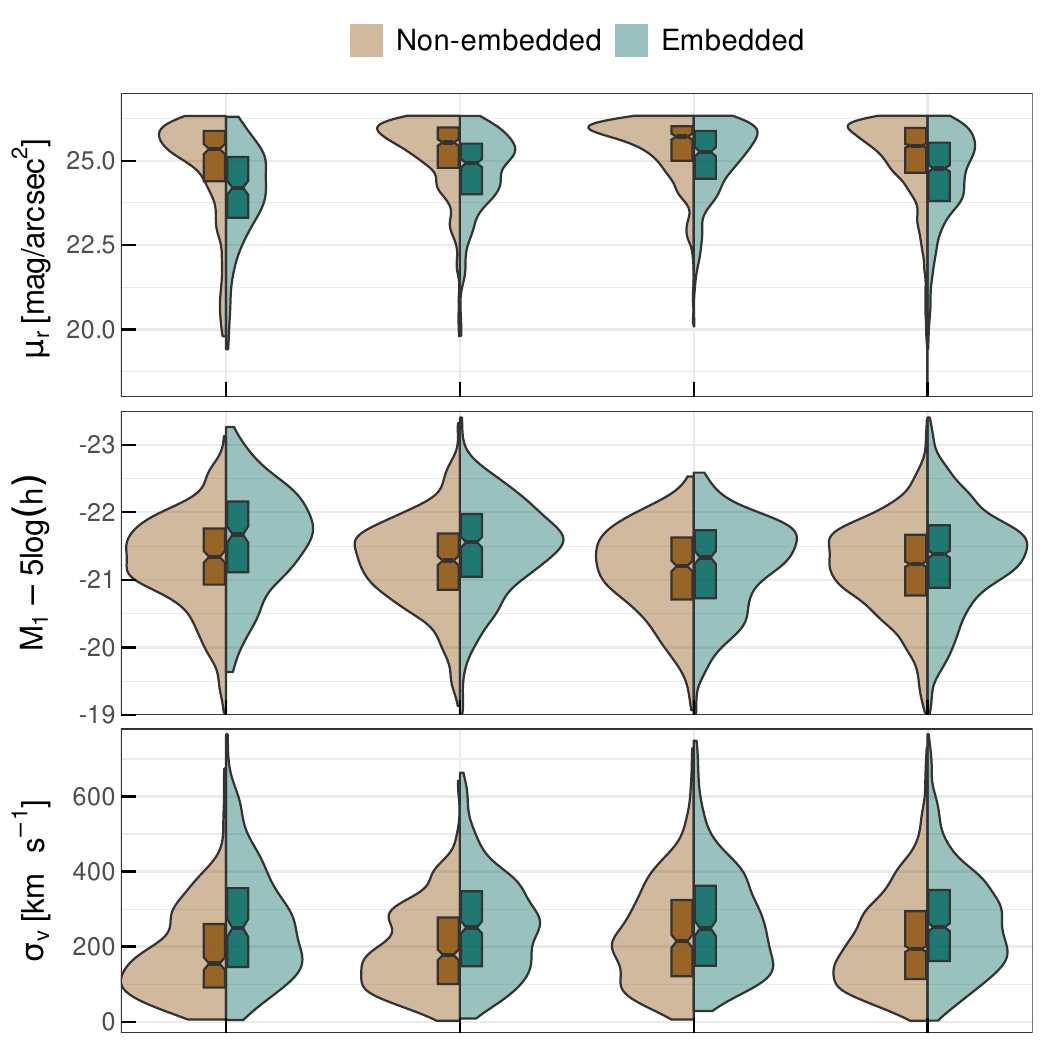}
   \includegraphics[width=0.49\textwidth]{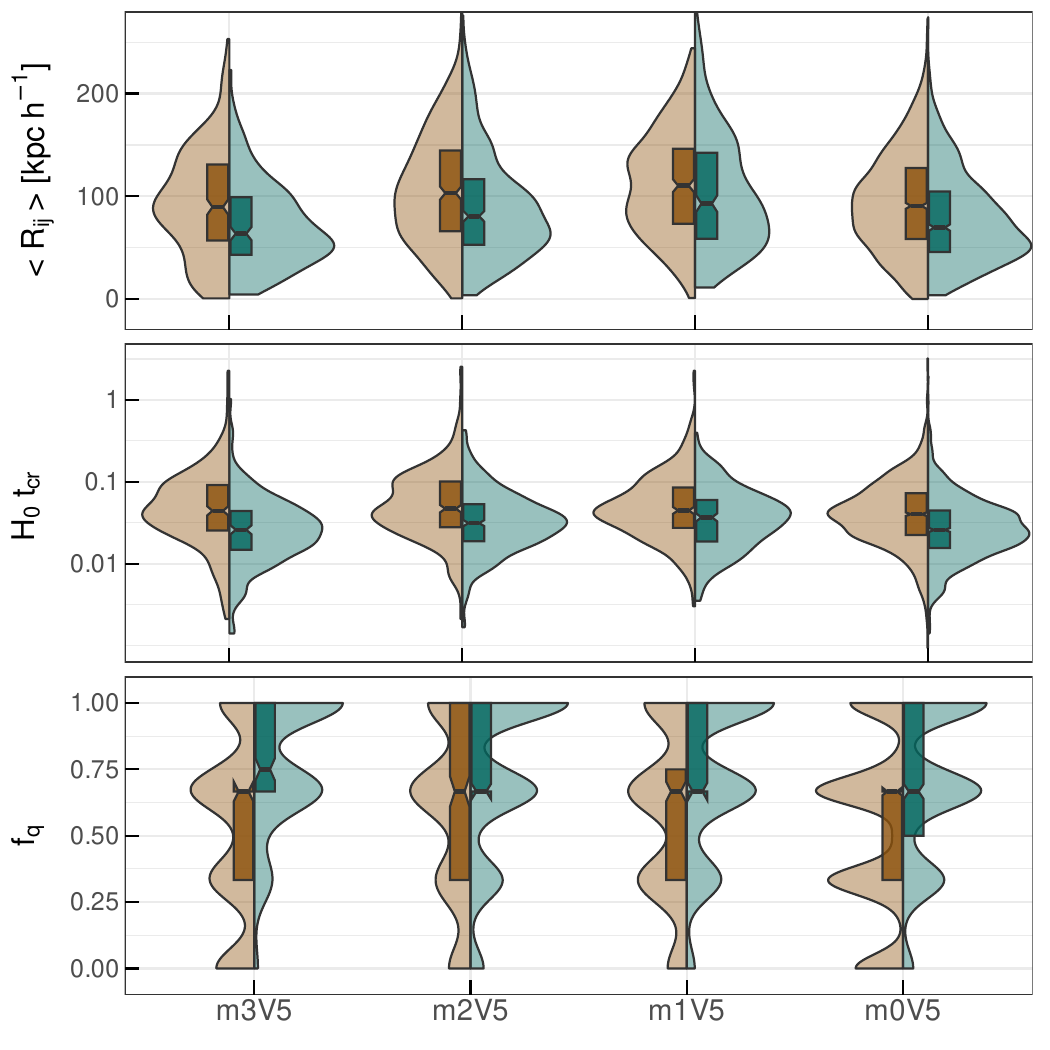}
      \caption{CG properties when the sample is split according to whether they can be considered embedded or not in a loose group.   
              }
         \label{fig:violin}
\end{figure}

\subsection{Galaxy quenching in CGs}
Early studies of \cite{Hickson92}, \cite{ribeiro+98} and \cite{pompei+12} have shown that there is a positive correlation in CGs between the fraction of spiral galaxies and the crossing time. CGs with the shortest crossing times display the lowest fraction of spiral galaxies, while those with the largest crossing times are prone to contain a larger fraction of spiral galaxies. 

These results can be interpreted in terms of star formation as a scenario where very dense environments, such as those present in very small CGs (with the lowest crossing times), are expected to have the lowest fraction of star-forming galaxies or, in other words, they are more efficient to suppress the star formation in galaxies than less dense environments. Therefore, we analyse the fraction of low star-forming galaxies in CGs in GAMA using two galaxy properties: the g-r galaxy colour and the specific star formation rate. 
The upper panels of Fig.~\ref{fig:6} show the median behaviour of galaxy colours and sSFR as a function of galaxy stellar mass for the samples of CGs in GAMA selected using $V_{lim}=500 \ km \ s^{-1}$ (green dots) as well as polynomial fittings to the scatter plots with the LOESS\footnote{\cite{loess}} (LOcally Estimated Scatterplot Smoothing) method (green curves) while the grey region is the 95\% confidence interval.
We also include the corresponding measurements for their control samples (red dots and curves). As expected, low-mass galaxies exhibit bluer colours and higher sSFR than high-mass galaxies. 
From the curves (and their confidence intervals) it can be seen that there is a small difference (except for the {\tt m0V5} sample) between CGs and control samples for galaxies with stellar masses lower than $10^{10} \ h^{-2} \ \mathcal{M}_{\odot}$ (the confidence intervals do not overlap). Low mass galaxies in CGs appear redder and with lower sSFR than their counterparts in control groups. These results are consistent with previous findings of \cite{montaguth+23} where lower median sSFR was observed in CGs than in control systems. 

Using the procedure described in Sect.~\ref{sec:samples_gama}, we selected red galaxies from the g-r galaxy colour and quenched galaxies using the sSFR and computed the fraction of red ($f_{\rm red}$) and quenched ($f_q$) galaxies in CGs. In the lower panels of Fig.~\ref{fig:6} we show these fractions as a function of the group dimensionless crossing times. 
The $f_{\rm red}$ of control groups show a peak near a crossing time of 0.05 and decrease towards the extremes for the {\tt m3V5, m2V5} and {\tt m0V5} subsamples. On the other hand, CGs exhibit relatively constant behaviour. At shorter crossing times, CGs have a larger fraction of red galaxies than control groups. 

On the other hand, $f_q$ for the control samples resembles the behaviour observed previously (for the $f_{\rm red}$), but CGs show a decreasing trend of the $f_q$ with crossing times for {\tt m3V5, m2V5, m1V5} and {\tt m0V5}. Hence, in this case, a clearer quenching of galaxies is observed for galaxies in CGs with shorter crossing times. 

Our results are consistent with those obtained by other authors studying the fraction of spiral galaxies as a function of crossing time in CGs \citep{Hickson92,ribeiro+98, pompei+12}. More recently, \cite{moura+20} found the same correlation of a low fraction of spiral galaxies in CGs with shorter crossing times. They also found, analysing early-type galaxies, that some mechanisms, such as tidal interactions, should be present in CGs that favour gas loss in their galaxy members. 
It has been suggested that the HI deficiency observed in CGs can be used as a hint for the stage of evolution of these systems. Tidal interactions are probably the main mechanisms to explain this HI deficiency observed in CGs, separating large amounts of gas in filamentary structures, or tails, as well as in a diffuse component \citep{verdes+01,borthakur+10,hess+17}. Hence, more evolved CGs are more likely to display a clear loss of gas content.
Also from the analysis of the spiral fraction with the crossing time for a sample of Hickson CGs, \cite{liu+22} found similar results. They obtained a relatively constant trend of the fraction of quiescent galaxies (defined from mid-infrared colours, \citealt{zucker+16}) with crossing times. In this work, we also obtain an almost constant behaviour of the fraction of red galaxies in CGs with crossing times, while the fraction of quenched galaxies shows a decreasing trend with crossing times. Our results align with those for the spiral fraction, suggesting that similar mechanisms likely account for both the decrease in the proportion of spiral galaxies and the increase in quenched galaxies in CGs with very short crossing times.

\subsection{Non-embedded and embedded CGs}
As stated in several works (e.g., \citealt{rood+94,mendel+11,DiazGimenez+15}), despite the isolation criterion for identifying CGs, these systems are only locally isolated since the criterion is performed relative to the CGs own size, which is quite small. 
Moreover, recent studies using synthetic catalogues have demonstrated that although approximately 60\% of the CGs identified using Hickson-like criteria are dense structures in 3D space, most of them are found inhabiting larger groups. This is because Hickson's isolation criterion does not effectively ensure global isolation in 3D real-space \citep{Taverna+22}.
Therefore, CGs identified using a Hickson-like algorithm could be embedded in the different galactic structures in the Universe. Recently, \cite{taverna+23} analysed statistically the location of Hickson-like CGs identified in the SDSS DR16 \citep{dr16} in the large-scale structure of the Universe and found that roughly 45\% of them can be considered embedded in loose groups, filaments and voids. Later, \cite{taverna+24} using several mock catalogues constructed from numerical simulations plus semianalytical models of galaxy formation confirmed that almost half of CGs are embedded and this percentage increases to roughly 60\% when considering CGs that have assembled early (more than 7 Gyrs ago).    

In this work, we follow \cite{taverna+23} and classify CGs as embedded or non-embedded in loose groups to analyse the main properties of CGs as a function of their location to other galactic structures. Briefly, using the sample of loose groups described in Sect.~\ref{sec:LGs}, we first remove the loose groups that are also identified as CGs\footnote{We consider a CG to be the same group as a loose group if they share at least 75\% of their members, and the number of bright loose group members that are beyond 3 times the CG radius is less than half of the CG members.}.
The percentage of loose groups considered equal to CGs are around 3-5\% when using as reference the samples {\tt m3V10}, {\tt m3V5}, {\tt m2V5}, and {\tt m1V5}, while 19\% of loose groups are selected as equal to CG when compared with {\tt m0V5} sample.
Then, using the remaining loose groups and the sample of CGs, we perform a member-to-member comparison and consider a CG embedded in a loose group if they share at least 2 members. Notice that these embedded CGs do not necessarily have all their members in common with the loose group. They may have extra members and yet we call them embedded CGs. 

The percentages of embedded CGs in GAMA are:  
42\% of the {\tt m3V10} and {\tt m3V5}, 49\% of the {\tt m2V5}, 45\% of the {\tt m1V5}, and 38\% of the {\tt m0V5}. In general, these percentages agree with the previously reported for CGs in SDSS \citep{zheng+20,taverna+23}. 

\begin{figure}
   \centering
   \includegraphics[trim={0 6cm 0 0}, clip, width=0.49\textwidth]{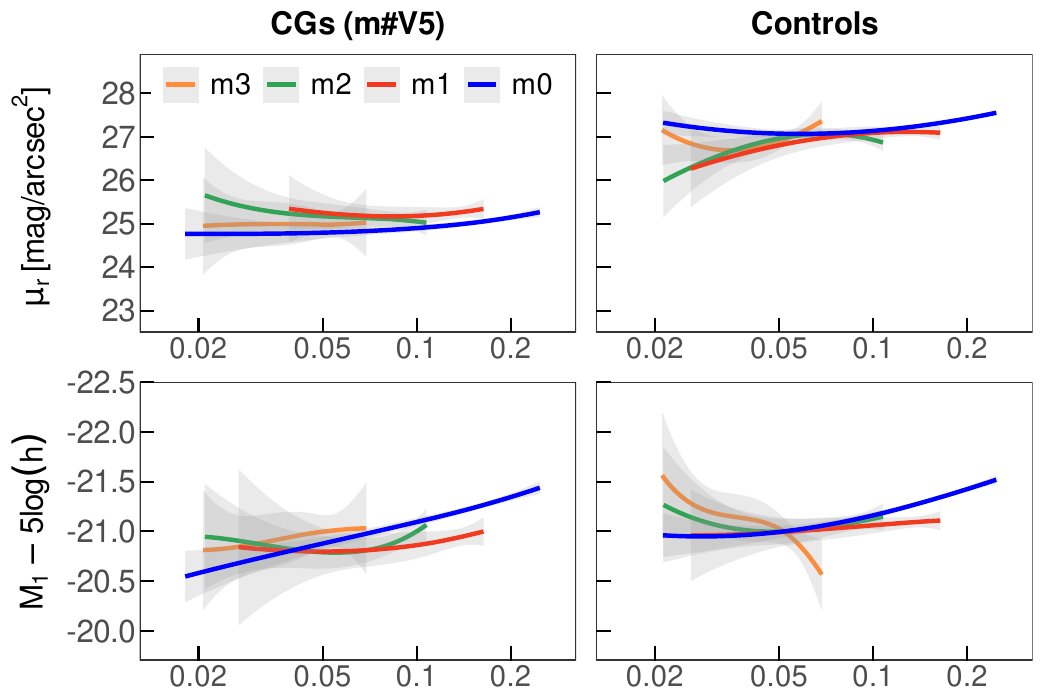}
   \includegraphics[trim={0 0.3cm 0 0.6cm}, clip, width=0.49\textwidth]{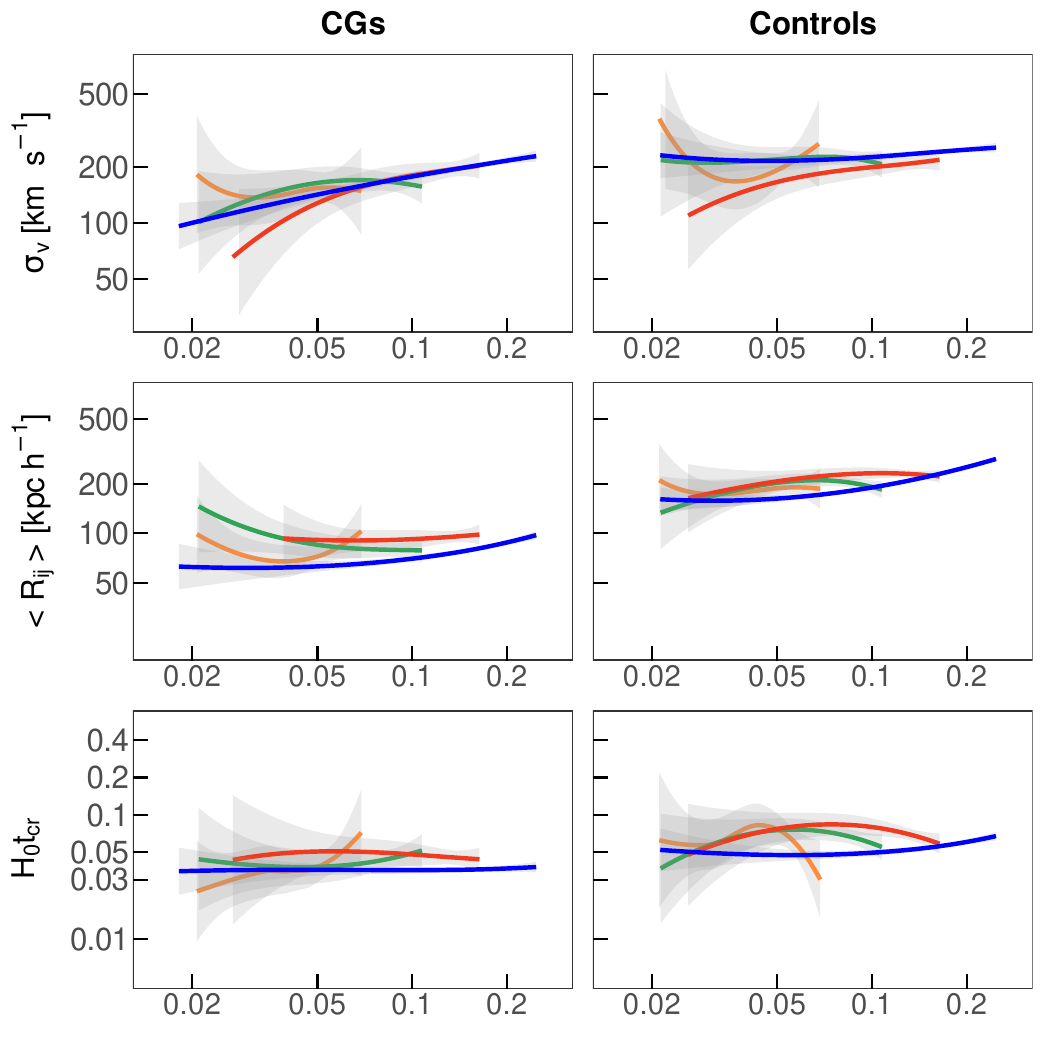}
   \includegraphics[trim={0 0.4cm 0 0.2cm}, clip, width=0.49\textwidth]{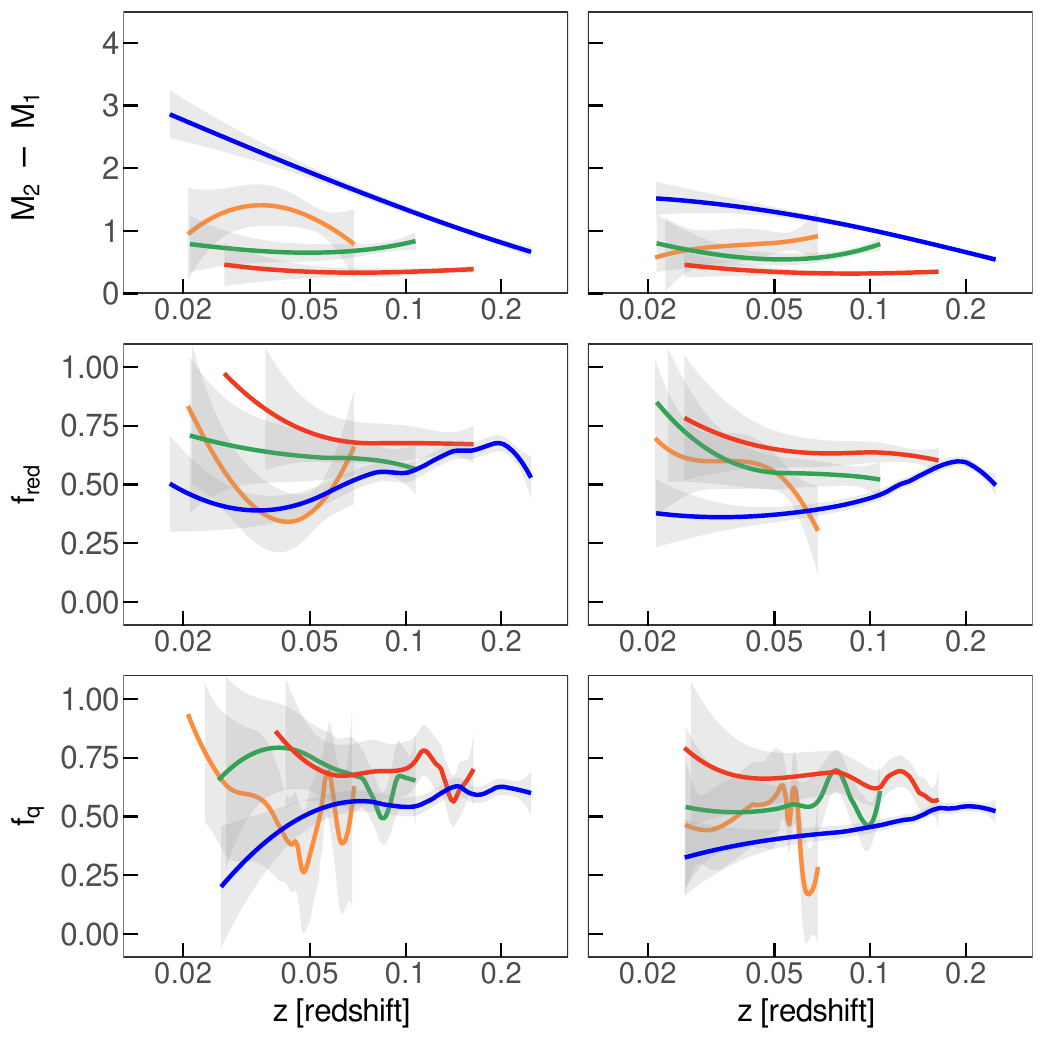}
      \caption{Group properties as a function of redshift for volume-limited samples. The left column is for CGs while the right column is for Control groups, both with $V_{lim}=500 \ km \ s^{-1}$. Curves are polynomial fittings to the scatter plots displaying roughly median values per redshift bin.
              }
         \label{fig:propz}
\end{figure}

It is worth reminding that varying $\Delta m$ in the CG searching also implies a variation in the isolation criterion. Therefore, {\tt m3V5}, {\tt m2V5} and {\tt m1V5} samples allow galaxies inside their isolation ring fainter than 3, 2 or 1 magnitudes, respectively, from the brightest galaxy of the CG. Hence, {\tt m1V5} is the most permissive of the three samples, which could mean that this sample might have more chances of being embedded in galaxy systems than the other two samples ({\tt m3V5} and {\tt m2V5}). However, according to the results, this is not the case. On the other hand, the {\tt m0V5} sample imposes the strongest restriction inside the isolation since no galaxy is allowed there. This could indicate that this sample might be less prone to be embedded. Our results might confirm this hypothesis since the {\tt m0V5} sample has between 4 to 11\% less embedded CGs than the other samples. In general, our results for CGs in GAMA agree with previous findings \citep{mendel+11, zheng+21,taverna+23,taverna+24}.

Figure~\ref{fig:violin} shows the violin diagrams 
of the distributions of properties of CGs split into non-embedded (left half) and embedded (right half) systems for compact groups identified with $V_{\rm lim}=500 \, \rm km/s$ and different magnitude gaps. From this figure, a clear distinction can be observed between embedded and non-embedded CGs. CGs considered embedded show the highest compactness with the smallest galaxy separations, the largest radial velocity dispersions, and the shortest crossing times, while also showing the brightest first-ranked galaxies and the largest fraction of quenched galaxies. These differences are more pronounced in the {\tt m3V5} sample, the most nearby redshift sample. The results are consistent with previous findings for these properties in \cite{taverna+23} and \cite{taverna+24}. Regarding the velocity dispersion, \cite{pompei+12} also found that most CGs relatively close to larger structures display large velocity dispersions which could imply some influence of the larger potential of the surrounding structures. 
More recently, \cite{montaguth+24} observed that the major structures that host CGs seem to accelerate the morphological transformation of CG galaxies. Therefore, they argue that the quenching of the CG members is due to interactions between galaxy members and also to interactions of the CG with its surroundings. 
Our results confirm and extend the previous findings to broader redshift ranges.

\subsection{CGs properties as a function of redshift}
\label{sec:zevol}
The depth of the GAMA catalogue enables building samples that span broader redshift ranges. In this section, we analyse the dependence of the CG properties with redshifts. To perform this analysis we build volume-limited samples of CGs (and control groups) for each sample identified with $V_{lim}=500 \ km \ s^{-1}$. This restriction facilitates the comparison of the properties of different redshift bins since we avoid the usual biases inherent in flux-limited samples. Therefore, we select CGs with the first-ranked galaxy absolute magnitude brighter than $-20$ in the r-band, and the group bi-weighted median redshift lower than 0.069, 0.1075, 0.164 and 0.248 for {\tt m3}, {\tt m2}, {\tt m1} and {\tt m0}, respectively (see vertical lines in bottom left panel of Fig.~\ref{fig:1}).  

Figure~\ref{fig:propz} shows the evolution of several CG properties as a function of redshift. From the analysis of these behaviours, most properties do not display a clear evolution with redshift. The samples {\tt m3V5}, {\tt m2V5} and {\tt m1V5} display an almost constant trend with redshift for the surface brightness, luminosity of the first-ranked galaxy, the median galaxy separation, and the magnitude gap. 
There is a small tendency for the {\tt m3V5} sample to show an increasing function for the crossing time of CGs with redshift, but the presence of large confidence intervals make this claim statistically unsupported.
Finally, we also observed an apparent trend (but not statistically reliable, Pearson coefficient $r=-0.19$ and $p=0.2$) of the {\tt m3V5} CGs to have a larger fraction of red and quenched galaxies at lower redshifts than that observed at larger distances. The {\tt m2V5} and {\tt m1V5} samples neither display a trend for the fraction of quenched galaxies with redshifts. Regarding {\tt m0V5} sample, some properties display some evolutionary trends with redshifts, such as the brightness of the first-ranked galaxy, the magnitude gap and the fraction of red and quenched galaxies. However, as we stated before, this sample is non-homogeneous because there is no magnitude restriction for the first-ranked galaxies (flux limit of the BGG criterion). This characteristic of the identification process can account for the redshift trends observed in these properties. 

From the comparison with the control groups, we observe that besides the general differences already mentioned previously (CGs are more compact, with smaller galaxy separation, radial velocity dispersion and crossing times, and larger magnitude gaps), the trends with redshift are similar to those observed for CGs. 

\cite{wilman+05} reported a decreasing trend for the fraction of red and quenched galaxies with redshifts in loose groups and field galaxies to at least up to $z\sim 0.45$, as well as a stronger effect on groups than in the field. They stated that the most likely scenario would be the presence of transforming mechanisms that account for the suppression of the star formation of the galaxies. In this work, we only find an apparent tendency (not statistical) for the m3V5 sample, the smallest and shallowest ($z_{cm}<0.069$) of our volume-limited samples.  A larger sample is needed to confirm these results.

\subsection{The Tremaine-Richstone statistics for CGs}
\label{sec:tr}

\begin{figure}
    \centering
    \includegraphics[width=0.5\textwidth]{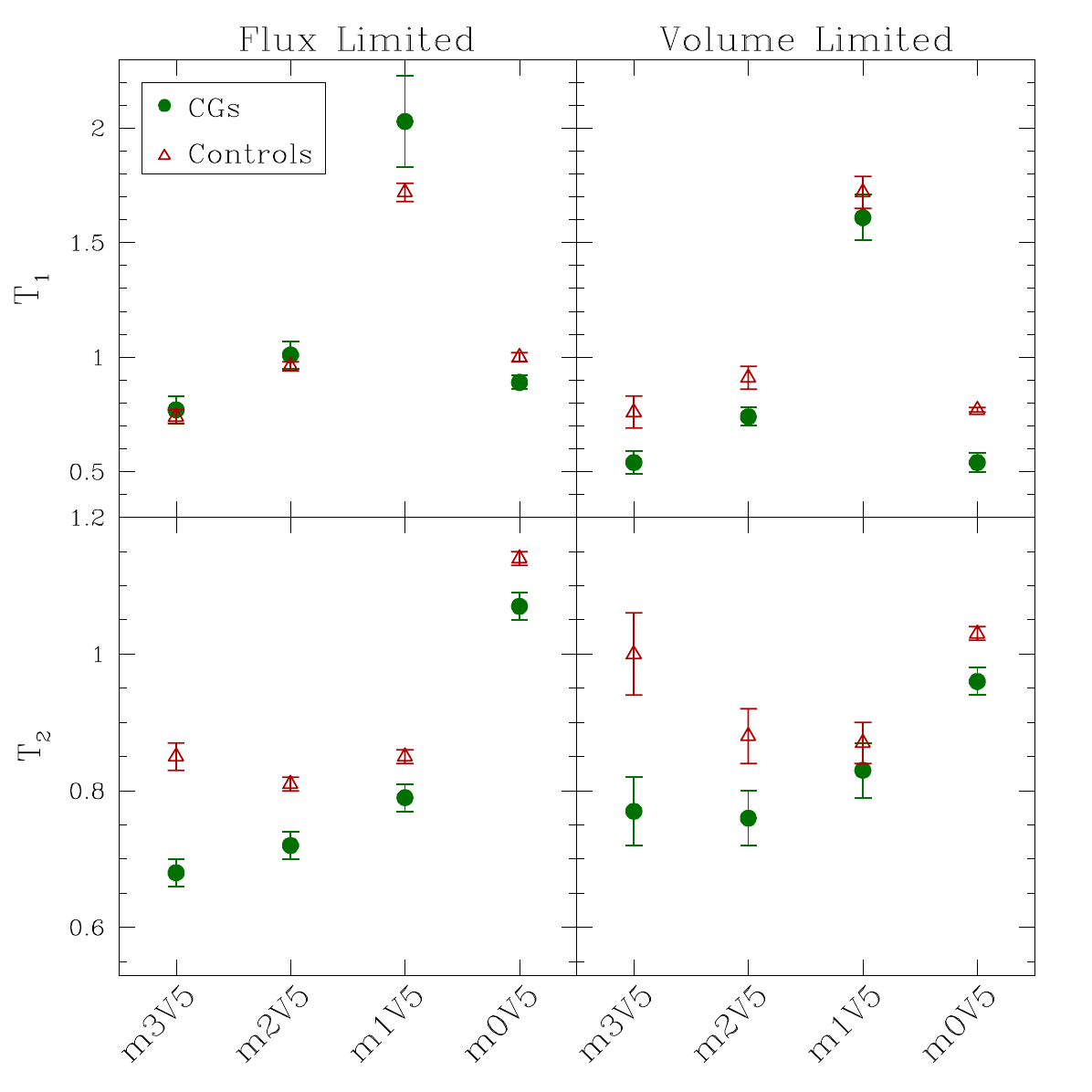}
    \caption{The Tremaine-Richstone statistics ${\rm T_1}$ and ${\rm T_2}$ \citep{tremaine+77} computed for the different samples of CGs (green circular points) and controls (red triangular points). Error bars are computed using the bootstrap technique. The left column displays the statistics computed for the flux-limited group samples while the right column displays the results for the volume-limited sample of groups.
    }
    \label{fig:t1t2}
\end{figure}
\cite{tremaine+77} proposed two powerful statistics to test whether the first-ranked galaxies in galaxy systems 
are consistent with a statistical sampling of a luminosity function. These statistics rely on the brightness of the first ranked galaxy ($M_1$) and the magnitude gap between the two brightest galaxies in the system ($M_2-M_1$). They are defined as follows:
$$T_1 = \frac{\sigma(M_1)}{\langle M_2-M_1 \rangle} \ \ \ \ \ {\rm and} \ \ \ \ \ T_2 = \frac{\sigma(M_2-M_1)}{\sqrt{0.677} \ \langle M_2-M_1 \rangle}$$
where $\sigma$ are the standard deviations, and angle brackets are the mean values. When $T_1$ and $T_2$ exceed unity, we can infer that the first-ranked galaxies are consistent with a random sampling of any given luminosity function. On the other hand, values of  $T_1$ and $T_2$ lower than unity suggest that the first-ranked galaxies are abnormally bright compared to their closest companion in luminosity. Galaxy interactions like galaxy mergers within systems could be responsible for the abnormal brightness of the main galaxy reducing the $T$ values below 0.7 \citep{Mamon87,mamon87TR}.

We compute the Tremaine-Richstone statistics for the flux- and volume-limited samples of CGs and control groups with $V_{lim}=500 \ km \ s^{-1}$. The results are shown in Fig.~\ref{fig:t1t2} with error bars computed from a bootstrap resampling technique. We show the results for the flux-limited samples as it is commonly used in the literature, however, a fairer comparison among the different samples ({\tt m3V5, m2V5, m1V5} and {\tt m0V5}) can be made when dealing with volume-limited samples. From the volume-limited samples, we observe that {\tt m3V5} and {\tt m2V5} CGs samples show $T_1$ values between 0.5-0.7 and $T_2$ around 0.75. These values are consistent with previous results, showing that first-ranked galaxies in CGs exceed the random sampling expectation. On the other hand, the larger than unity value of $T_1$ obtained for the {\tt m1V5} CGs sample indicates that, despite their compactness, these systems are not as evolved as the {\tt m3V5} and {\tt m2V5} CGs samples. This result is expected since, by construction, {\tt m1V5} CG sample comprises systems whose galaxy members have very similar luminosities among them. In addition, despite their inhomogeneities, {\tt m0V5} CG sample displays a very low value for $T_1$ ($\sim 0.5$), however, $T_2$ value is barely below unity. Finally, in general, the control samples display larger values for $T_1$ and $T_2$ than the observed for CGs, with the most notorious difference observed in $T_2$ when a gap of 3 magnitudes is considered, obtaining a value consistent with unity. Therefore, when considering the luminosity of their two brightest galaxies, the differences observed in $T_1$ and $T_2$ suggest that CGs may exhibit signs of more evolved systems compared to loose groups.

\section{Summary and Conclusions}
\label{sec:conclusions}
In this work, we present a series of CG catalogues extracted from the GAMA redshift survey. We identified CG following the guidelines of the Hickson criteria but modifying a set of parameters to produce different CG samples:
\begin{itemize}
    \item we adopt a limiting galaxy velocity difference in the line-of-sight of 500 km $s^{-1}$. This choice reduces by a factor of two the usual limit of 1000 km $s^{-1}$ adopted in several previous works. Our value is selected to reduce misidentification by spuriously linking galaxies in the radial direction.
    \item we vary the allowed magnitude range between the brightest and the faintest galaxy members. We produce different CG samples by setting this magnitude range in 3, 2 and 1 magnitudes, plus a sample where no magnitude range is introduced (i.e., every galaxy is allowed to be classified as a group member). It should be noted, that these choices impact 3 of the 5 criteria that define the Hickson-like procedure (population, flux limit of the BGG and isolation).
\end{itemize}

These choices in the searching algorithm parameters allowed us to build CG samples with varying depths, thus probing the intermediate redshift range. 
The samples with a given magnitude range (3, 2 or 1) comprise between $500$ and $700$ systems with median redshifts between $0.1$ and $0.18$, and maximum redshifts of $0.2$ to $0.32$. The sample without a restriction in the magnitude range of its galaxy members comprises more than $2\, 000$ systems with a median redshift of $0.16$ and a maximum redshift of $0.43$.
We performed several tests on these samples to analyse the CG properties and their galaxy members. We also build control samples for each CG sample from the catalogue of loose groups in the parent galaxy survey identified following the procedure of \cite{rodriguez2020combining}. 
As a summary, we can highlight the following results:
\begin{itemize}
    \item Analyzing the properties of each CG sample we observe that, in general, most of the properties behave similarly between samples with different sets of parameters (except for those properties that are directly dependent on the identification parameters). In particular, we observe a small brightening of the first-ranked galaxy and a decreasing projected separation between galaxies when moving from 1 to 3 in the magnitude range. These results are consistent with previous CG samples \citep{zandivarez+22}. We also observe that the sample of CGs with no magnitude restriction displays similar properties to those of CGs obtained by  \cite{zheng+20} identified with similar restrictions.
    \item Control samples of loose groups tailored to reproduce the membership magnitude range and the velocity restrictions of CGs show lower compactness, larger galaxy separations and higher velocity dispersions than the observed in the corresponding CG counterparts. 
    \item The fraction of quenched galaxies in CGs is higher than in their control samples. These differences are mainly observed for galaxies with low stellar masses ($10^{10} \ h^{-2} \mathcal{M}_{\odot} $) that inhabit CGs with the shortest crossing times. 
    It has been shown that low-mass galaxies in compact groups have practically lost all their hot gas content while keeping a small reservoir of cold gas at present, which strongly influences the suppression of the star formation rate \citep{zandivarez+23}.
    This result resembles the previous findings about the fraction of spiral galaxies with the crossing times in CGs \citep{Hickson92,ribeiro+98,pompei+12,moura+20}.
    \item We also analyse the influence of the near surroundings of CGs by splitting the samples according to their condition of embedded or not in a loose galaxy system. We obtain that on average 44\% of CGs are embedded in loose groups. A similar percentage of embedded CGs was previously reported by \cite{taverna+23} for CGs in the SDSS and by \cite{taverna+24} in CGs identified in several mock catalogues built from different semianalytical models of galaxy formation. Our study also confirms previous findings that show that embedded CGs display the highest compactness and velocity dispersions, shorter crossing times and brightest first-ranked galaxies \citep{taverna+23}. In addition, the distribution of the fraction of quenched galaxies is shifted toward larger values for embedded compact groups, in agreement with the findings of \cite{montaguth+24}. These behaviours are consistent with CGs characterised by an early assembly\footnote{See \cite{DiazGimenez+21} for a detailed description of assembly channels in Hikcson-like CGs.} \citep{zandivarez+23} which dominate the sample of embedded CGs \citep{taverna+24}. 
    \item Given the depths reached for our CG samples, we investigate the possible variation of CG properties as a function of redshift. To perform a fair comparison, we built volume-limited samples with an absolute magnitude limit of $-20$ in the r-band. We do not find evolution of CG properties with redshift for the CG samples selected with magnitude ranges of 1, 2, and 3. However, several properties display some evolution with redshift for the sample of CGs built without a magnitude restriction. Nevertheless, these behaviours are likely to be caused by the intrinsic inhomogeneities inherent in this sample.  
    \item Finally, we complement our analysis by computing the Tremaine-Richstone statistics of the CG samples \citep{tremaine+77}. Our estimate of the $T_1$ and $T_2$ values shows signs that the first-ranked galaxies of the CGs are anomalously bright. This result is clear for the CG samples with a membership magnitude range of 2 and 3 magnitudes. 
    As expected, the CG sample with members within a 1-magnitude range shows different values than the other CG samples (mainly for $T_1$) because galaxy members in these systems have very similar luminosities among each other by construction. These results that support the abnormally bright first-ranked galaxy in CGs in GAMA are in agreement with previous estimates obtained for other observational \citep{DiazGimenez+12} and synthetic CGs catalogues \citep{Taverna+16}. It has been previously suggested that the abnormal brightness of the main galaxy may be attributed to the dynamic interactions, such as the merging of galaxies within its system \citep{ostriker+77,Mamon87,mamon87TR,ostriker+19}.
\end{itemize}

Our findings for compact groups in GAMA support the scenario in which these highly dense systems tend to harbour exceptionally luminous first-ranked galaxies, likely due to interactions and/or mergers with their companions. This scenario makes compact groups more favourable places for the suppression of the star formation rate than loose systems. This phenomenon is enhanced in those compact groups that have formed earlier and are most likely embedded in larger systems. Ultimately, the extreme proximity of galaxies and their nearby environment conspire to accelerate the quenching of their galaxies.

As a corollary, our work presents a new set of statistically reliable compact group catalogues with high redshift completeness that manage to sample the intermediate redshift zone and will be available for the astronomical community. These catalogues might represent the beginning of reliable statistical studies of these extreme systems beyond the local Universe, and will serve as a benchmark for CG studies in upcoming deep spectroscopic surveys such as DESI and others.

\begin{acknowledgements}
We thank the anonymous referee for their suggestions that improved the final version of the manuscript.
This publication uses as a parent catalogue the GAMA survey. GAMA is a joint European-Australasian project based around a spectroscopic campaign using the Anglo-Australian Telescope. The GAMA input catalogue is based on data taken from the Sloan Digital Sky Survey and the UKIRT Infrared Deep Sky Survey. Complementary imaging of the GAMA regions is being obtained by a number of independent survey programmes including GALEX MIS, VST KiDS, VISTA VIKING, WISE, Herschel-ATLAS, GMRT and ASKAP providing UV to radio coverage. GAMA is funded by the STFC (UK), the ARC (Australia), the AAO, and the participating institutions. The GAMA website is \url{http://www.gama-survey.org/}.\\

This publication also uses the SDSS Data Release 16 (DR16) which is one of the latest data releases of the SDSS-IV.
Funding for the Sloan Digital Sky 
Survey IV has been provided by the 
Alfred P. Sloan Foundation, the U.S. 
Department of Energy Office of 
Science, and the Participating 
Institutions. SDSS-IV acknowledges support and resources from the Center for High 
Performance Computing  at the 
University of Utah. The SDSS 
website is \url{www.sdss.org}.
SDSS-IV is managed by the 
Astrophysical Research Consortium 
for the Participating Institutions 
of the SDSS Collaboration including 
the Brazilian Participation Group, 
the Carnegie Institution for Science, 
Carnegie Mellon University, Center for 
Astrophysics | Harvard \& 
Smithsonian, the Chilean Participation 
Group, the French Participation Group, 
Instituto de Astrof\'isica de 
Canarias, The Johns Hopkins 
University, Kavli Institute for the 
Physics and Mathematics of the 
Universe (IPMU) / University of 
Tokyo, the Korean Participation Group, 
Lawrence Berkeley National Laboratory, 
Leibniz Institut f\"ur Astrophysik 
Potsdam (AIP),  Max-Planck-Institut 
f\"ur Astronomie (MPIA Heidelberg), 
Max-Planck-Institut f\"ur 
Astrophysik (MPA Garching), 
Max-Planck-Institut f\"ur 
Extraterrestrische Physik (MPE), 
National Astronomical Observatories of 
China, New Mexico State University, 
New York University, University of 
Notre Dame, Observat\'ario 
Nacional / MCTI, The Ohio State 
University, Pennsylvania State 
University, Shanghai 
Astronomical Observatory, United 
Kingdom Participation Group, 
Universidad Nacional Aut\'onoma 
de M\'exico, University of Arizona, 
University of Colorado Boulder, 
University of Oxford, University of 
Portsmouth, University of Utah, 
University of Virginia, University 
of Washington, University of 
Wisconsin, Vanderbilt University, 
and Yale University.\\

This work has been partially supported by Consejo Nacional de Investigaciones Científicas y Técnicas de la República Argentina (CONICET) and the Secretaría de Ciencia y Tecnología de la Universidad de Córdoba (SeCyT).\\

FR would like to acknowledge support from the
ICTP through the Junior Associates Programme 2023-2028
\end{acknowledgements}

\bibliography{refs} 

\appendix

\section{Catalogues of compact groups and loose groups in the GAMA survey}
\label{app_t}

In this appendix, we display sample tables of the compact and loose groups catalogues derived from the Galaxy and Mass Assembly spectroscopic survey \citep{gama, gama2}. 

We present the first catalogues of compact groups derived from the GAMA survey. These catalogues encompass five samples of CGs: {\tt m3V10}, {\tt m3V5}, {\tt m2V5}, {\tt m1V5}, and {\tt m0V5} (see Table \ref{tab:samples}).
Table \ref{tab:groups} details the properties of the CGs included in each catalogue, and Table \ref{tab:members} shows the properties of the compact group members. Given the similarity in the format of the tables for each sample, we only show an example from the {\tt m3V10} catalogue.

\begin{table*}
\caption{Compact groups identified in GAMA (example from sample {\tt m3V10}). \label{tab:groups}}
\begin{center}
\tabcolsep=3.5pt
\begin{tabular}{ccccccccccccc}
\hline
\hline
\texttt{CG}id & $N$ & RA & Dec  & Redshift & $\langle R_{ij} \rangle$ & $\mu_r$  & $\sigma_v$ & $H_0 \, t_{cr}$ & $r_{\rm b}$ &$r_{\rm f}$ & $M_{1}-5log(h)$ & $M_{2}-5log(h)$  \\
 & & [deg] & [deg] & &   [${\rm kpc \, h^{-1}}$] & [$\rm mag \, arsec^{-2}$]  & [${\rm km \ s^{-1}}$] & & [mag] & [mag] & [mag] & [mag]\\
\hline
1&	3	&178.323 &	1.211	& 0.079026	& 56.646 &	23.974 & 271.924 &	0.019 &	15.628 &	17.342 &	-21.459 &	-20.765\\
2&	3	&183.002 &	-0.769	& 0.073621	& 11.840 &	21.661 & 268.045 &	0.004 &	16.659 &	17.664 &	-20.238 & 	-19.993\\
3&	3	&220.771 &	-0.656	& 0.153248	& 74.146 &	23.697 & 529.314 &	0.013 &	16.677 & 	18.732 &	-21.867 &	-20.951\\
4&	3	&179.979 &	-0.323	& 0.084799	& 45.505 & 	23.139 & 154.031 &	0.027 &	15.630 &	18.304 &	-21.606 &	-19.457\\
5&	3	&183.136 &	0.176	& 0.083562	& 214.525 &	26.166 & 54.604	& 0.356	 &15.978 &	16.601 &	-21.172 &	-20.704\\
\vdots & \vdots & \vdots & \vdots & \vdots & \vdots & \vdots & \vdots & \vdots & \vdots & \vdots & \vdots & \vdots \\
\hline
\end{tabular}
\end{center}
\parbox{\hsize}{Notes: \texttt{CG}id: compact group ID, $N$: number of galaxy members, RA: group centre right ascension (J2000), Dec: group centre declination (J2000), $Redshift$: group CMB redshift,  $\langle R_{ij} \rangle$: median of the projected separations between galaxy members, $\mu_r$: r-band group surface brightness, $\sigma_v$: radial velocity dispersion, $H_0 \, t_{cr}$: dimensionless crossing time, $r_{\rm b}$ / $r_{\rm f}$: r-band observer-frame model apparent magnitude of the brightest/faintest galaxy, $M_1-5log(h)$ / $M_2-5log(h)$: rest-frame absolute magnitude of the brightest/second brightest galaxy of the group. This table is available in electronic form.}
\end{table*}

\begin{table}
\caption{Galaxy members of compact groups identified in GAMA (example from sample {\tt m3V10}).\label{tab:members}}
\begin{center}
\tabcolsep=2.5pt
\begin{tabular}{ccccccc}
\hline
\hline
\texttt{CG}id & RA & Dec & Redshift  & $r$ & $g$ & $M_r - 5log(h)$ \\
  & [deg] & [deg] &  & [mag] & [mag]& [mag]\\
\hline
1& 	178.316& 	1.211& 	0.079545& 	15.628 & 16.573 & -21.459\\
1& 	178.315& 	1.205& 	0.079162& 	16.304 & 17.153 & -20.765\\
1& 	178.330& 	1.218& 	0.078292& 	17.342 & 18.231 & -19.704\\
2& 	183.001& 	-0.771& 	0.073618& 	16.659 & 17.458 & -20.238\\
2& 	183.003& 	-0.769& 	0.073624& 	16.910 & 17.794 & -19.993\\
2& 	183.003& 	-0.766& 	0.072388& 	17.664 & 18.355 & -19.185\\
 \vdots & \vdots & \vdots & \vdots & \vdots & \vdots & \vdots\\
\hline
\end{tabular}
\end{center}
\parbox{\hsize}{Notes: \texttt{CG}id: compact group ID, RA: right ascension (J2000), Dec: declination (J2000), Redshift: CMB redshift, $r$ and $g$: r-band and g-band observer-frame model apparent magnitudes corrected for extinction in the AB system; $M_r-5 log(h)$: rest-frame absolute magnitude in the r-band. Galaxies of each group are ordered by their apparent magnitudes from brightest to faintest.
This table is available in electronic form.}
\end{table}


We also include the header of the catalogue of loose groups identified in the GAMA survey. Table \ref{tab:fof} shows the properties of the groups and Table \ref{tab:fof_members} lists the properties of the galaxy members.

The full version of the whole set of catalogues used in this work will be available using the VizieR catalogue access tool\footnote{CDS, Strasbourg Astronomical Observatory, France, DOI: \url{10.26093/cds/vizier}} \citep{vizier}.

\begin{table}
\caption{Loose groups identified in GAMA. \label{tab:fof}}
    \begin{center}
    \tabcolsep=3.5pt
        \begin{tabular}{ccccccc}
        \hline
        \hline
        \texttt{LG}id & $N$ & RA & Dec  & Redshift & Mass & $r_b$ \\
         & & [deg] & [deg] & &   [$\mathcal{M}_\odot \, h^{-2}$] & [mag]\\
        \hline
        1 &	2 &	174.023 &	0.705 &	0.332885 &	1.846e+13 &	18.534 \\
        2 &	2 &	174.101 &	0.660 &	0.230415 &	3.770e+12 &	18.771 \\
        3 &	2 &	174.110 &	0.809 &	0.328046 &	1.172e+13 &	19.128 \\
        \vdots & \vdots & \vdots & \vdots & \vdots & \vdots & \vdots \\
        \hline
        \end{tabular}
    \end{center}
    \parbox{\hsize}{Notes. \texttt{LG}id: loose group ID, $N$: number of galaxy members, RA: group centre right ascension (J2000), Dec: group centre declination (J2000), $Redshift$: group CMB redshift, Mass: group mass, $r_{\rm b}$: r-band observer-frame model apparent magnitude of the group brightest galaxy. This table is available in electronic form.}
\end{table}

\begin{table}
\caption{Galaxy members of loose groups identified in GAMA.\label{tab:fof_members}}
\begin{center}
\tabcolsep=2.5pt
\begin{tabular}{ccccccc}
\hline
\hline
\texttt{LG}id & RA & Dec & Redshift & $r$ & $g$ & $M_r - 5log(h)$ \\ 
 & [deg] & [deg] & & [mag] & [mag]& [mag]\\
\hline
1 &	174.022 &	0.703 &	0.333610	& 19.304 &	20.634 &-21.433\\
1 &	174.023 &	0.706 &	0.332460	& 18.534 &	19.609 &-22.014\\
2 &	174.101 &	0.659 &	0.230580	& 18.771 &	19.995 &-20.960\\
2 &	174.100 &	0.661 &	0.230110	& 19.414 &	20.602 &-20.293\\
3 &	174.109 &	0.804 &	0.327519	& 19.128 &	20.542 &-21.612\\
3 &	174.111 &	0.817 &	0.328789	& 19.573 &	21.100 &-21.241\\
 \vdots & \vdots & \vdots & \vdots & \vdots  & \vdots & \vdots \\
\hline
\end{tabular}
\end{center}
\parbox{\hsize}{Notes: \texttt{LG}id: loose group ID, RA: right ascension (J2000), Dec: declination (J2000), Redshift: CMB redshift, $r$ and $g$: r-band and g-band observer-frame model apparent magnitudes corrected for extinction in the AB system; $M_r-5 log(h)$: rest-frame absolute magnitude in the r-band. This table is available in electronic form.}
\end{table}

\end{document}